\documentclass[twocolumn]{aastex631}

\usepackage{printlen}
\usepackage{amssymb}
\usepackage{amsmath}
\usepackage{amsfonts}
\usepackage{color}

\newcommand{\jetopen}{b}
\newcommand{\jetdist}{d}
\newcommand{\tesc}{t_{\rm ej}{}}

\newcommand{\zeltron}{\textsc{zeltron}}
\DeclareMathOperator{\Li}{Li}
\makeatletter
\newcommand\thefontsize{\f@size pt}
\makeatother

\graphicspath{{./}}

\begin{document}

\title{Radiative relativistic turbulence as an \textit{in situ} pair-plasma source in blazar jets}

\author[0000-0002-7414-0175]{John Mehlhaff}
\affiliation{Univ.\ Grenoble Alpes, CNRS, IPAG, 38000 Grenoble, France; {\normalfont \url{john.mehlhaff@univ-grenoble-alpes.fr}}}

\author[0000-0003-0709-7848]{Muni Zhou}
\affiliation{Department of Physics and Astronomy, Dartmouth College, Hanover, NH 03755, USA}
\affiliation{School of Natural Sciences, Institute for Advanced Study, Princeton, NJ 08544, USA}

\author[0000-0003-3816-7896]{Vladimir Zhdankin}
\affiliation{Department of Physics, University of Wisconsin-Madison, Madison, WI 53706, USA}

\begin{abstract}

As powerful gamma-ray engines, blazars -- relativistic plasma jets launched toward Earth from active galactic nuclei -- are manifestly high-energy particle accelerators. Yet, exactly how these jets accelerate particles as well as what they are made of both remain largely mysterious. In this work, we argue that these issues may be linked through the gamma-ray emission for which blazars are renowned. Namely, high-energy photons produced at sites of intense particle acceleration could be absorbed by soft radiation within the jet, enriching it with electron-positron pairs. We explore this possibility in the specific context of particle acceleration by magnetized radiative relativistic turbulence. Using a combination of theory, particle-in-cell simulations, and Fokker-Planck modeling, we identify and characterize a novel pair-production-mediated equilibration mechanism in such turbulence. Initially, turbulent energy injection outpaces radiative cooling, leading to runaway particle acceleration and gamma-ray radiation. Then, gamma-ray absorption begets copious newborn pairs, slowing subsequent particle acceleration. This eventually brings particle acceleration into balance with radiative cooling and shuts down pair production: a pair-enriched final equilibrium. We estimate that this process could significantly load jets of flat-spectrum radio quasars with fresh pairs. These results represent an important connection between particle acceleration and plasma composition in blazar jets. 

\end{abstract}


\section{Introduction}
\label{sec:intro}
Active galactic nuclei can launch powerful relativistic jets, that, when pointed towards Earth, are observed as blazars. Blazars shine across the electromagnetic spectrum, with spectra generally showing two broad nonthermal humps \citep{fossati_etal_1998, madejski_sikora_2016, blandford_etal_2019, prandini_ghisellini_2022}.
The lower-energy hump -- which extends from the radio up through the UV and, sometimes, X-ray bands -- is generally accepted to arise from synchrotron radiation by relativistic particles in the jet \citep{burbridge_1956, marscher_1980}.
Meanwhile, the higher-energy component, peaked in the gamma rays, is often thought to be produced by inverse Compton (IC) scattering of soft radiation by jet electrons and (if present) positrons (\citealt{jones_etal_1974, konigl_1981, ghisellini_etal_1985, begelman_sikora_1987, dermer_etal_1992}; though hadronic models for the gamma-ray component also exist: \citealt{mannheim_biermann_1992, aharonian_2000}).

Despite this rich phenomenology concerning the main radiative processes at play, what the jet plasma is made of and how the energetic particles responsible for the observed radiation are accelerated remain prominent outstanding questions \citep{madejski_sikora_2016, blandford_etal_2019}. In this work, we posit that these two issues may be intimately connected, as indeed the broadband aspect of blazar spectra suggests. For example, 10-GeV gamma rays, near the highest energies observed from blazars by the \textit{Fermi} satellite, may be absorbed by UV photons (of energy~$m_e^2 c^4/10\, \rm GeV \sim 30 \, \rm eV$), which lie in the lower-energy spectral hump and, hence, are probably present in the jet. Gamma-ray emission sites could thus alter the plasma composition \textit{in situ} by producing electron-positron pairs \citep{blandford_levinson_1995}.

These considerations are particularly poignant when applied to the most powerful blazar subclass: the flat-spectrum radio quasars (FSRQs). FSRQs show strong optical and UV emission lines. These suggest the presence both of a luminous accretion disk and of circumnuclear material. The latter intercepts and reprocesses the disk emission, redirecting onto the jet some of the radiated energy from the disk that would have otherwise missed the jet \citep{ghisellini_tavecchio_2008, madejski_sikora_2016}. FSRQ jets thus initially plow through an intense inner radiative environment, one rich with target photons not only for IC scattering, but also for absorbing IC-scattered gamma-rays.

Let  us make these ideas more concrete. At the parsec scale, the ambient radiation in FSRQ jets is probably dominated by that impinging from the circumnuclear broad-line and hot-dust regions (BLR and HDR), which, unlike the direct emission from the disk at these distances, is strongly blueshifted into the jet-comoving frame \citep{sikora_etal_1994, sikora_etal_2009, blandford_levinson_1995, ghisellini_tavecchio_2009, nalewajko_etal_2014, costamante_etal_2018, dmytriiev_etal_2025}. Given that the BLR and HDR shine, respectively, UV and IR radiation onto the jet, they become pair-production opaque at respective energies~$(m_e c^2)^2 / 10\, \mathrm{eV} \sim 30 \, \rm GeV$ and~$(m_e c^2)^2 / 0.1\, \mathrm{eV} \sim 3 \, \rm TeV$. Thus, the same particle acceleration processes that power observed GeV and TeV emission in FSRQs may also, at just slightly higher (and at least partially absorbed) photon energies, enrich the jet with electron-positron pairs.

The main objective of this work is to explore this possibility in the context of relativistic turbulence, a potential particle acceleration mechanism in blazar jets. Such jets are expected to be launched highly magnetized \citep{blandford_znajek_1977, blandford_payne_1982}. In this situation, relativistically magnetized turbulence -- potentially triggered by kink
\citep{
begelman_1998,             
tomimatsu_etal_2001,       
mizuno_etal_2009,          
oneill_etal_2012,          
alves_etal_2018,           
bodo_etal_2019,            
davelaar_etal_2020,        
ortuno-macias_etal_2022,   
musso_etal_2024}           
or shear-driven
\citep{
turland_scheuer_1976,     
fiedler_jones_1984,       
hardee_etal_2007,         
hamlin_newman_2013,       
sironi_etal_2021,         
chow_etal_2023,           
cerutti_giacinti_2023,    
davelaar_etal_2023,       
figueiredo_etal_2024,     
tsung_etal_2025}          
instabilities in the jet -- provides an efficient mechanism for depleting the magnetic free energy to power particle acceleration, gamma-ray emission and, potentially, pair production
\citep{
zhdankin_etal_2017,
zhdankin_etal_2018,
zhdankin_etal_2020,
zhdankin_etal_2021,
uzdensky_2018,
comisso_sironi_2018,
comisso_sironi_2019,
comisso_sironi_2021,
sobacchi_etal_2021a,
sobacchi_etal_2021b,
nattila_beloborodov_2021,
hankla_etal_2022,
vega_etal_2022b,
vega_etal_2022a,
bresci_etal_2022,
davis_etal_2022,
davis_etal_2024,
groselj_etal_2024,
imbrogno_etal_2024,
singh_etal_2025,
nattila_2024}.
As a candidate particle acceleration process, turbulence does not preclude magnetic reconnection
\citep[also commonly invoked in blazar studies;][]{
giannios_etal_2009,
nalewajko_etal_2011,
giannios_2013,
sironi_etal_2015,
werner_etal_2018,
christie_etal_2019,
mehlhaff_etal_2021,
mehlhaff_etal_2024}, since reconnecting current sheets may form at the dissipation scales of the turbulent cascade \citep[][]{zhdankin_etal_2013}, while turbulence itself may be triggered by reconnection \citep{kowal_etal_2017, guo_etal_2021}.

Turbulence is already a complex multiscale problem, involving nontrivial interaction between the plasma microscales and the global dynamics. This cross-scale coupling is only further enriched by strong radiative cooling and pair production, which may load the turbulent zone with fresh plasma, feeding back on subsequent particle acceleration (as recently highlighted by \citealt{groselj_etal_2024} and \citealt{nattila_2024}). This feedback is therefore a central theme of the present study. To isolate its intrinsic aspects, we first construct a local, but nevertheless FSRQ-inspired, model of relativistic radiative turbulence. Afterwards, we employ the main quantities output by our local model to set it back in a global context via analytic estimates. We find that, for generous but potentially attainable jet magnetizations, radiative turbulence could significantly load FSRQ jets with newborn pairs.

In particular, in Section~\ref{sec:sketch}, we theoretically predict the salient features of pair-regulated turbulence in FSRQ jets. Then, in Section~\ref{sec:pic}, we employ first-principles particle-in-cell (PIC) simulations to validate our theoretical model. Subsequently, in Section~\ref{sec:fp}, we use our PIC results to construct and benchmark a simplified numerical model in which we assume that turbulent particle acceleration is diffusive, obeying a Fokker-Planck (FP) equation. Simulating the evolution of this PIC-inspired FP equation is much cheaper than the PIC simulations of Section~\ref{sec:pic}, and therefore enables a more thorough and quantitative exploration of the parameter space of pair-regulated FSRQ turbulence. The FP models of Section~\ref{sec:fp} furnish the main quantitative results that, in Section~\ref{sec:astroapps}, we feed into analytic estimates to gauge the extent to which turbulence may enrich FSRQ jets with fresh pairs. We conclude in Section~\ref{sec:conclusions}.

\section{Theoretical model}
\label{sec:sketch}
We consider magnetized relativistic pair-plasma turbulence. Homogeneous turbulence is continuously driven by injection of energy into fluctuations at an outer length scale~$L$. The mean magnetic field and initial electron+positron number densities are, respectively,~$B_0\boldsymbol{\hat{z}}$ and~$n_0$. We specialize to the plasma producing high-energy radiation, assuming ultrarelativistic particle Lorentz factors,~$\gamma \equiv (1-v^2/c^2)^{-1/2} \gg 1$, where $v$ is the particle velocity. The characteristic plasma magnetization is~$\sigma = B_0^2 / (4 \pi w)$ where~$w = 4 \langle \gamma \rangle m_e c^2 n_0 / 3$ is the plasma enthalpy density and~$\langle \gamma \rangle \gg 1$ is the system-averaged Lorentz factor. The magnetization controls the Alfvén speed,~$v_A = c [\sigma / (1 + \sigma)]^{1/2}$.

To mimic the photons impinging on the jet from the BLR or the HDR, we immerse the plasma in a soft ambient radiation bath of total energy density~$U_{\rm rad}$. We take this radiation to be homogeneous, isotropic, and monochromatically distributed at energy~$\epsilon_{\rm rad} \ll m_e c^2$. The photons comprising the bath are targets for IC scattering by energized plasma particles. When a photon is IC-scattered to energy above~$(m_e c^2)^2/\epsilon_{\rm rad}$, it can subsequently be absorbed by the background to produce an electron-positron pair. 

\subsection{Radiative preliminaries}
\label{sec:primer}

The photon energy~$\epsilon_{\rm rad}$ can be recast in terms of the critical Lorentz factor~$\gamma_{\rm KN} = m_e c^2 / (4 \epsilon_{\rm rad})$ \citep{blumenthal_gould_1970, mehlhaff_etal_2021}. A particle with~$\gamma < \gamma_{\rm KN}$ IC-scatters background photons continuously in the Thomson regime. For~$\gamma > \gamma_{\rm KN}$, on the other hand, IC scattering enters the quantum-electrodynamic (QED) Klein-Nishina limit: each scattered photon carries away an order-unity fraction of the scattering particle's energy.

The mean IC power radiated by a single particle is
\begin{align}
    P_{\rm IC}(\gamma) = P_{\rm T}(\gamma) f_{\rm KN}(\gamma / \gamma_{\rm KN}) \, ,
\end{align}
where~$f_{\rm KN}(x)$ is the dimensionless function \citep{jones_1968}
\begin{align}
    f_{\rm KN}(x) &= \frac{9}{x^3} \left[ \left( \frac{x}{2} + 6 + \frac{6}{x} \right) \ln \left( 1 + x \right) - 2 + 2 \Li_{2}(-x) \right. \notag \\
    &- \left. \frac{1}{\left(1 + x\right)^2} \left( \frac{11}{12}x^3 + 6 x^2 + 9 x + 4 \right) \right] \,
    \label{eq:fkn}
\end{align}
and~$\Li_{2}$ is the dilogarithm.
\citet{moderski_etal_2005} found the approximate expression
\begin{align}
    f_{\rm KN}(x) \simeq \frac{1}{(1+x)^{3/2}} \, ,
    \label{eq:fknapprox}
\end{align}
valid to within a factor of~$3$ until~$x$ exceeds roughly~$10^4$. 
In the limit~$\gamma \ll \gamma_{\rm KN}$,~$f_{\rm KN}$ tends toward unity, and, hence,~$P_{\rm IC}(\gamma)$ simplifies to the Thomson radiated power,
\begin{align}
    P_{\rm T}(\gamma) = \frac{4}{3} c \sigma_{\rm T} \gamma^2 U_{\rm rad} \, ,
    \label{eq:thomsonpower}
\end{align}
where~$\sigma_{\rm T} = 6.6 \times 10^{25} \, \rm cm^{2}$ is the Thomson cross section.

The IC radiative cooling time of a particle is
\begin{align}
    t_{\rm cool}(\gamma) \equiv \frac{\gamma m_e c^2}{P_{\rm IC}(\gamma)} = \frac{L}{c} \frac{\gamma_{\rm cool}}{\gamma} \frac{1}{f_{\rm KN}(\gamma/\gamma_{\rm KN})} \, ,
    \label{eq:tcool}
\end{align}
where we have defined the Lorentz factor~$\gamma_{\rm cool} = 3 m_e c^2 / (4 \sigma_{\rm T} U_{\rm rad} L)$ as that of a particle which, in the Thomson~($\gamma \ll \gamma_{\rm KN}$) limit, cools down in one lightcrossing time of the driving scale, $L/c$. Due to the presence of~$f_{\rm KN}(\gamma/\gamma_{\rm KN})$ in~(\ref{eq:tcool}),~$t_{\rm cool}(\gamma)$ is non-monotonic. For,~$\gamma \ll \gamma_{\rm KN}$, it decreases with~$\gamma$, but, for~$\gamma \gg \gamma_{\rm KN}$, it increases. In this way, Klein-Nishina effects cause~$t_{\rm cool}$ to acquire a global minimum of~$ct_{\rm cool,min}/L \simeq 2.3 \gamma_{\rm cool} / \gamma_{\rm KN}$ at~$\gamma \simeq 3.2 \gamma_{\rm KN}$ \citep{mehlhaff_etal_2021}.

Besides transitioning to a quantum radiative cooling regime, particles with~$\gamma > \gamma_{\rm KN}$ also tend to scatter photons to above pair-production threshold,~$(m_e c^2)^2 / \epsilon_{\rm rad}$, with the background radiation.\footnote{Technically, a particle needs Lorentz factor~$4 \gamma_{\rm KN}$ (instead of~$\gamma_{\rm KN}$) to be able to IC scatter photons to above pair-production threshold. However, we will mostly ignore this factor of~$4$.} The pair-production cross section,~$\sigma_{\gamma\gamma}$, quickly rises from~$0$ to about~$\sigma_{\rm T} / 5$ for scattered photon energies between~$(m_e c^2)^2 / \epsilon_{\rm rad}$ and~$3.6 (m_e c^2)^2 / \epsilon_{\rm rad}$, after which point it declines slowly with increasing energy. We therefore use the peak value,~$\sigma_{\gamma\gamma} \simeq \sigma_{\rm T} / 5$, to define a characteristic optical depth~$\tau_{\gamma \gamma} = U_{\rm rad} \sigma_{\rm T} L / (5 \epsilon_{\rm rad})$ to pair production presented by the background radiation across the driving scale~$L$.

Using our definitions for~$\gamma_{\rm KN}$ and~$\gamma_{\rm cool}$, one can write~$\tau_{\gamma\gamma} = 3 \gamma_{\rm KN} / (5 \gamma_{\rm cool})$. Thus, the typical lifetime of an above-threshold photon,~$L / (c \tau_{\gamma\gamma})$, is also roughly the minimum-possible cooling time,~$t_{\rm cool,min} \simeq 2.3 L \gamma_{\rm cool} / (c \gamma_{\rm KN}) \simeq 1.4 L/ (c \tau_{\gamma\gamma})$, of a radiating particle. It is a critical timescale introduced by QED physics. 

While~$\tau_{\gamma\gamma}$ likely exceeds unity (Section~\ref{sec:astroapps}), the optical depth of the turbulent plasma to Thomson scattering,~$\tau_{\rm T} = n_0 \sigma_{\rm T} L$, remains small (i.e.,~$n_0 \ll 1 / (\sigma_{\rm T} L) \ll U_{\rm rad} / \epsilon_{\rm rad}$). Most IC seed photons that traverse the turbulent region thus pass through unaffected. That is, IC scattering and pair production do not feed back on the ambient radiation. However, the few lucky seed photons that do get scattered are likely absorbed soon thereafter: while still inside the turbulent plasma.

\subsection{Previous Thomson-regime results}
\label{sec:prevthom}
A regime of radiative turbulence close to that described above has already been studied by \citet{zhdankin_etal_2020}. Those authors specialized to Thomson~($\gamma \ll \gamma_{\rm KN}$) IC cooling without pair production. They found that turbulence in this case relaxes to a quasisteady state characterized by a thermal (modulo intermittent fluctuations) particle energy distribution. The steady-state (normalized) plasma temperature,~$\theta = kT/(m_e c^2)$, adjusts so that the rate of turbulent energy injection per particle,
\begin{align}
    \dot{\mathcal{E}}_{\rm inj} = \eta \frac{B_0^2}{8 \pi n_0} \frac{v_A}{L},
    \label{eq:einj}
\end{align}
balances the per-particle radiated power,
\begin{align}
    \dot{\mathcal{E}}_{\rm rad} = \langle P_{\rm T} \rangle = \frac{4}{3} c \sigma_{\rm T} \langle \gamma^2 \rangle U_{\rm rad} = 16 c \sigma_{\rm T} \theta^2 U_{\rm rad} \, .
    \label{eq:eradthom}
\end{align}
In the above,~$\eta \sim 1$ is an efficiency factor; angle brackets denote averaging over the turbulent region;~$\dot{\mathcal{E}}_{\rm inj}$ is defined by assuming that the turbulent energy density $\langle \delta \boldsymbol{B}^2 \rangle/(8\pi)$ (with fluctuating magnetic field~$|\delta \boldsymbol{B}| \sim B_0$) dissipates over a cascade time $\sim L/v_A$; and, for an ultrarelativistic Maxwell-Jüttner (i.e., thermal) distribution,~$\langle \gamma^2 \rangle = 12 \theta^2$. Equating~(\ref{eq:einj}) to~(\ref{eq:eradthom}) results in a steady-state temperature,
\begin{align}
    \theta_{\rm ss} = \frac{\eta}{6} \frac{v_A}{c} \sigma \gamma_{\rm cool} = \frac{\eta}{6} \sqrt{\frac{\sigma^3}{1+\sigma}} \gamma_{\rm cool} \, .
    \label{eq:thss}
\end{align}
This is an implicit equation for~$\theta_{\rm ss}$, since the magnetization~$\sigma$ also depends on~$\theta_{\rm ss}$ through the enthalpy density,~$w = 4 \langle \gamma \rangle n_0 m_e c^2 / 3 = 4 \theta_{\rm ss} n_0 m_e c^2$.

\citet{zhdankin_etal_2020} pointed out that a quasithermal steady state is consistent with the hypothesis that particle acceleration is diffusive \citep[as, indeed, had already been shown in PIC simulations by][]{wong_etal_2020}. To retrace their argument, we introduce the particle energy distribution function,~$f(\gamma)$, defined such that~$f(\gamma) d \gamma$ is the number of particles with Lorentz factor between~$\gamma$ and~$\gamma + d \gamma$. A thermal distribution can be retrieved if~$f(\gamma)$ obeys the FP equation,
\begin{align}
     \partial_t f = \partial_\gamma \left( D \partial_\gamma f \right) - \partial_\gamma \left[ \left( 2 D / \gamma + A \right) f \right] \, ,
     \label{eq:fp}
\end{align}
with diffusion coefficient~$D(\gamma) = \gamma^2 / t_{\rm acc}$ at high energies \citep[as found by][]{wong_etal_2020} and advection coefficient~$A(\gamma)=A_{\rm T}(\gamma)=-\gamma^2 c / (L \gamma_{\rm cool})$ corresponding to Thomson radiative cooling (see also earlier work by \citealt{schlickeiser_1985}). In that case, solving~(\ref{eq:fp}) in the steady state yields
\begin{align}
    f(\gamma) \propto \gamma^2 \exp \left( -\frac{c t_{\rm acc}}{L \gamma_{\rm cool}} \gamma \right) \, ,
    \label{eq:fpsoln}
\end{align}
an ultrarelativistic Maxwell-Jüttner distribution with temperature~$\theta = L \gamma_{\rm cool} / (c t_{\rm acc})$. This temperature is equal to that,~$\theta_{\rm ss}$, predicted by~(\ref{eq:thss}) provided that
\begin{align}
    \frac{L}{c t_{\rm acc}} = \frac{\eta}{6} \sqrt{\frac{\sigma^3}{1+\sigma}} \, .
    \label{eq:taccconstraint}
\end{align}
We note that, due to the cancellation of~$\gamma_{\rm cool}$, no radiative parameter~($U_{\rm rad}$ or~$\epsilon_{\rm rad}$) appears in equation~(\ref{eq:taccconstraint}), suggesting that it is a condition on the consistency of the diffusive particle acceleration ansatz independently of radiative cooling.
Indeed, the scaling of~$D(\gamma)$ with~$\sigma$ predicted by~(\ref{eq:taccconstraint}) agrees with that recently measured from non-radiative PIC simulations by \citet{wong_etal_2025}.

\subsection{Expectations for the Klein-Nishina IC regime}
\label{sec:knexpectations}
We here make some basic predictions for the regime where~$\gamma$ may exceed~$\gamma_{\rm KN}$. First, we assume that particle acceleration remains diffusive even if cooling transitions to a different regime. Let us see what this implies for the steady-state solution of equation~(\ref{eq:fp}), ignoring pair production for the moment. Plugging in the same diffusion coefficient as before,~$D(\gamma) = \gamma^2 / t_{\rm acc}$, but modifying the advection coefficient to~$A_{\rm IC}(\gamma) = A_{\rm T}(\gamma) f_{\rm KN}(\gamma / \gamma_{\rm KN}) = -\gamma^2 c f_{\rm KN}(\gamma / \gamma_{\rm KN}) / (L \gamma_{\rm cool})$ to crudely\footnote{A proper treatment requires accounting for discrete (non-infinitesimal) radiative energy loss, which cannot be captured by an advection coefficient. This is done in Sections~\ref{sec:pic} and~\ref{sec:fp}.} incorporate Klein-Nishina effects yields the solution
\begin{align}
    f(\gamma) \propto \gamma^2 \exp \left( 2 \frac{c t_{\rm acc} \gamma_{\rm KN}}{L \gamma_{\rm cool}} \frac{1}{\sqrt{1 + \gamma/\gamma_{\rm KN}}} \right) \, .
    \label{eq:fpsolnkn}
\end{align}
Here, we have used approximation~(\ref{eq:fknapprox}) to obtain a closed form, but using the exact~$f_{\rm KN}$ does not change the following main conclusion. Namely, the distribution~(\ref{eq:fpsolnkn}) is not normalizable. Instead, because radiative cooling is less efficient~($f_{\rm KN}(x) < 1$) in the Klein-Nishina regime, it is no longer able to keep up with particle acceleration. We therefore expect particle acceleration to run away once it breaks past~$\gamma=\gamma_{\rm KN}$.

The existence of pair production does not change this conclusion. If a pair is produced inside the turbulent zone, then the absorbed high-energy photon did not leave the system. Thus, pair creation only further dampens the collective plasma radiative efficiency, retaining energy that would have otherwise been radiated away.

In addition, as long as pair production persists, a steady state cannot be reached; the overall number of pairs just keeps increasing. Therefore, the only way for a steady state to be achieved is if the turbulence can self-regulate so as to quench pair production. A main point of this paper is that, given enough time, such self-regulation is the inevitable fate of relativistic turbulence coupled to a seed-photon background.

Equation~(\ref{eq:taccconstraint}) provides a hint as to how such regulation might occur. According to that equation, the efficiency of particle acceleration is tied to the plasma magnetization,~$\sigma$. The magnetization, in turn, decreases with increasing pair density. This suggests that a steady state may be achieved if enough pairs are generated to drive down~$\sigma$ and cut particle acceleration off before pair production threshold: at Lorentz factors less than~$\gamma_{\rm KN}$. In such a situation, radiative cooling relaxes to the Thomson IC regime (since~$\gamma < \gamma_{\rm KN}$). We know the form of the steady-state particle energy distribution in this case: it is thermal, as given by~(\ref{eq:fpsoln}).

We thus envision the following sequence of events:
\begin{enumerate}
    \item Turbulence-induced nonthermal particle acceleration is initially very efficient, yielding many particles with Lorentz factors~$\gamma > \gamma_{\rm KN}$ that emit pair-producing gamma rays.
    \item Over time, newborn pairs begin to the load the plasma magnetization,~$\sigma$, inducing a corresponding drop in particle acceleration efficiency.
    \item No more pairs are produced once particle acceleration cuts off below~$\gamma_{\rm KN}$. The plasma then thermalizes with temperature~$\theta_{\rm ss} < \gamma_{\rm KN}$ set, as in equation~(\ref{eq:thss}), by the final pair-loaded magnetization.
\end{enumerate}
In the following section, we present PIC simulations to show that this sequence of events indeed transpires.

\section{PIC-simulation proof of concept}
\label{sec:pic}
We present~3D PIC simulations of driven turbulence using the \zeltron~code \citep{cerutti_etal_2013, cerutti_werner_2019} modified, as detailed by \citet{mehlhaff_etal_2024}, to incorporate the QED effects of interest here. Our simulation setup follows closely that detailed by \citet{zhdankin_etal_2018}. We employ a cubic periodic box of volume~$L^3$. The box is initially threaded by a uniform magnetic field~$\boldsymbol{B}_0 = B_0 \boldsymbol{\hat{z}}$ and filled with a homogeneous electron+positron number density~$n_0$ sampled from a Maxwell-Jüttner distribution of temperature~$\theta_{0} = 100$. We choose~$B_0$ and~$n_0$ such that the initial magnetization is~$\sigma_{\rm 0} = B_0^2 / (16 \pi n_0 \theta_0 m_e c^2) = 2.5$. The box size and magnetic field strength together define the system-size-limited Lorentz factor,~$\gamma_{\rm max} = e B_0 L / (m_e c^2)$: that of a particle whose gyroradius equals~$L$.

Unlike in the setup of \citet{zhdankin_etal_2018}, we resolve the starting number density~$n_0$ with just one electron and one positron per cell. This allows room for the plasma number density to grow several fold over the course of the simulation without exhausting computer memory. On the other hand, it also increases the initial particle noise, which we mitigate by employing 10 digital current filter passes per timestep.

We resolve the box size,~$L$, with~$N=512$ grid cells in each dimension. As a convergence check, we have also performed a simulation with~$N=768$ and otherwise identical parameters. Smaller simulations did not resolve all the important plasma scales (discussed below). Although resource constraints prevent us from integrating the~$N=768$ simulation for as long as the~$N=512$ run, which is our primary focus, the two are identical in their early-time overlap.

We substantially overresolve the initial Debye length,~$\lambda_{\rm 0} = [\theta_0 m_e c^2 / (4 \pi n_0 e^2)]^{1/2}$, with~$8$ grid cells~($12$ cells for the~$N=768$ run). This allows the Debye length to shrink as the particle count grows and the plasma cools without triggering spurious numerical heating. Our system-size-limited Lorentz factor is set to~$\gamma_{\rm max} = 2 \times 10^4$.

For the radiative parameters, we choose~$U_{\rm rad}$ such that the initial temperature,~$\theta_{\rm 0}$, corresponds to the steady-state temperature that the plasma would keep, according to equation~(\ref{eq:thss}) with~$\eta = 1$, if Klein-Nishina effects were absent. This corresponds to~$\gamma_{\rm cool} \simeq 300$. We then choose~$\gamma_{\rm KN} = 500$ such that the fiducial optical depth~$\tau_{\gamma\gamma} = 3 \gamma_{\rm KN} / (5 \gamma_{\rm cool}) \simeq 1$. This value of~$\gamma_{\rm KN}$ also means that all particles start below pair-production threshold. Hence, all pairs born \textit{in situ} are the result of self-consistent turbulent particle acceleration rather than spuriously energetic initial particles.

Starting at time~$t=0$, we drive turbulence by injecting an external randomly fluctuating current density into the Maxwell-Ampère Law \citep{tenbarge_etal_2014}. This excites magnetic field fluctuations~$\delta \boldsymbol{B}$ comparable in strength to the initial field:~$\langle \delta \boldsymbol{B}^2 \rangle \simeq B_0^2$. To achieve convincing convergence of macroscopic quantities toward a radiative steady state, we integrate our simulation for more than~$200$ lightcrossing times of the box (we pushed the~$N=768$ run through just over~$100$ lightcrossing times).

\begin{figure}
    \centering
    \includegraphics[width=\columnwidth]{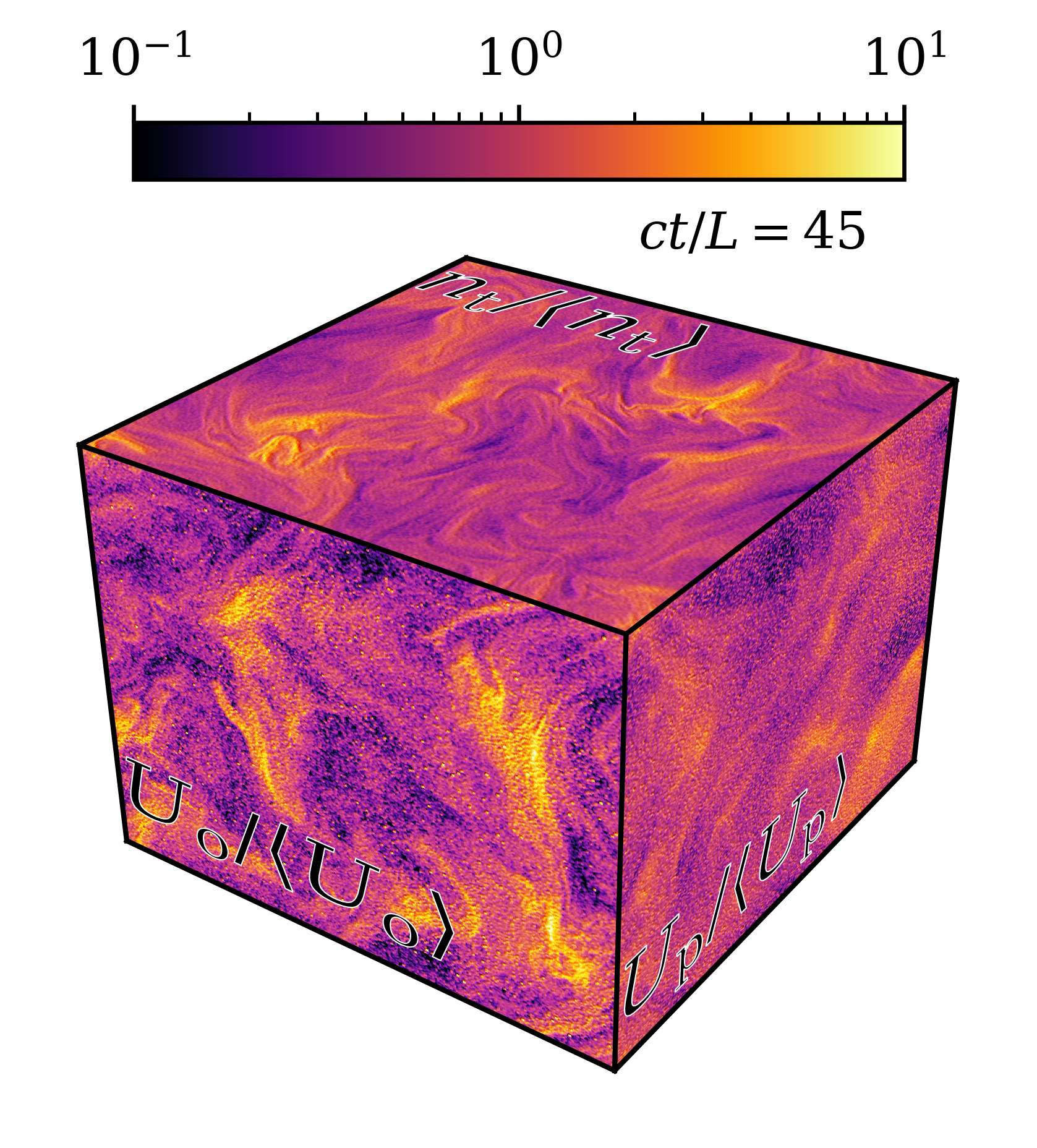}
    \caption{Snapshot of the total plasma number density,~$n_{\rm t}$, energy density of original particles,~$U_{\rm o}$, and energy density of produced particles,~$U_{\rm p}$, each normalized by its instantaneous spatial average. The time chosen, $ct/L = 45$, is during the intermediate evolution.
    }
    \label{fig:box}
\end{figure}

\begin{figure}
    \centering
    \includegraphics[width=\columnwidth]{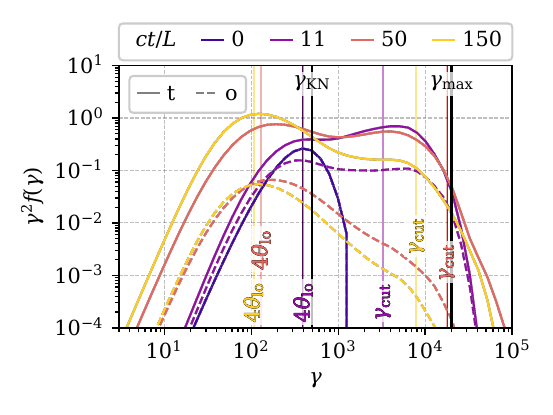}
    \caption{Snapshots of the total (t) particle energy distribution and that (o) of original particles only. Except for~$t=0$, the total distribution is dominated by the produced particles at all times shown. The~$\gamma_{\rm cut}$ and~$\theta_{\rm lo}$ values are measured using~(\ref{eq:gcut}) and~(\ref{eq:thetalodef}), respectively.}
    \label{fig:pspecs}
\end{figure}

We present a still of our simulation spatial domain in Fig.~\ref{fig:box} and several snapshots of the particle energy distribution function in Fig.~\ref{fig:pspecs}. Fig.~\ref{fig:box} shows the plasma number and energy densities at an intermediate time, close to the~$ct/L=50$ curve of Fig.~\ref{fig:pspecs}. At this point, the produced particles (those born on-the-fly as the result of pair production) completely dominate the original particles (those placed by hand at time~$t=0$), both in terms of overall particle count and plasma energy. Moreover, from the~$ct/L=50$ curve in Fig.~\ref{fig:pspecs}, we see that the produced particles contain about an equal amount of plasma energy at low Lorentz factors as at high ones (where gyroradii approach the system size).

The distributions in Fig.~\ref{fig:pspecs} illustrate the main phases of our simulation as anticipated in Section~\ref{sec:knexpectations}. At early times (the~$ct/L=11$ curve), turbulence drives strong nonthermal particle acceleration that initially outcompetes IC cooling. This results in a long tail in the particle distribution that extends almost up to the system-size limit,~$\gamma_{\rm max}$. Next, at intermediate times (the~$ct/L=50$ curve), a low-energy thermal bump emerges in addition to the high-energy tail. This bump has a temperature~$\theta_{\rm lo} < \gamma_{\rm KN}$: its particles do not emit pair-producing photons. Finally (the~$ct/L=150$ curve), the nonthermal tail declines, leaving predominantly the low-energy quasithermal bump. This quasithermal distribution is the one predicted in Section~\ref{sec:knexpectations}. Extrapolating to even later times, we would expect the nonthermal tail to eventually decay entirely, leaving solely the predicted quasithermal distribution at low energies.

We measure the cutoff Lorentz factor,~$\gamma_{\rm cut}$, in the particle energy distribution,~$f(\gamma)$, as \citep[cf.][]{sironi_etal_2016}
\begin{align}
    \gamma_{\rm cut} = \frac{d}{ds} \frac{\langle \gamma^{s+1} \rangle}{\langle \gamma^s \rangle}
    \label{eq:gcut}
\end{align}
where we set~$s=4$ and where
\begin{align}
    \langle \gamma^s \rangle = \frac{\int d \gamma \, \gamma^s f(\gamma)}{\int d \gamma \, f(\gamma)} \, .
    \label{eq:avggam}
\end{align}
Equation~(\ref{eq:gcut}) corresponds to the exponential cutoff Lorentz factor~$\gamma_{\rm cut}$ assuming~$f(\gamma) \propto \gamma^{-l} \exp(-\gamma/ \gamma_{\rm cut})$ for some power~$l<s$.
Our~$\gamma_{\rm cut}$ measurements are displayed in Fig.~\ref{fig:pspecs}. They are comparable to, but slightly lower than, the system limit, with~$\gamma_{\rm cut} \lesssim \gamma_{\rm max}$.

We also measure the temperature,~$\theta_{\rm lo}$, of the low-energy thermal hump in the particle energy distribution,~$f(\gamma)$. We define
\begin{align}
    \theta_{\rm lo} = \left[ \frac{2 \langle \gamma^p \rangle}{\Gamma(p+3)} \right]^{1/p} \, ,
    \label{eq:thetalodef}
\end{align}
where~$\Gamma(x)$ is the gamma function and~$p$ is empirically chosen (it must be larger than~$-2$). Formally,~(\ref{eq:thetalodef}) corresponds to the temperature one would measure if~$f(\gamma)$ were proportional to the Maxwell-Jüttner distribution,~$\gamma^2 \exp(-\gamma / \theta_{\rm lo})$. We choose a low exponent~$p=-1.8$, which permits a reasonable estimate for the temperature of the low-energy thermal hump by beating down the contribution from the ever-present (yet, at late times, subdominant) nonthermal tail.

\begin{figure}
    \centering
    \includegraphics[width=\columnwidth]{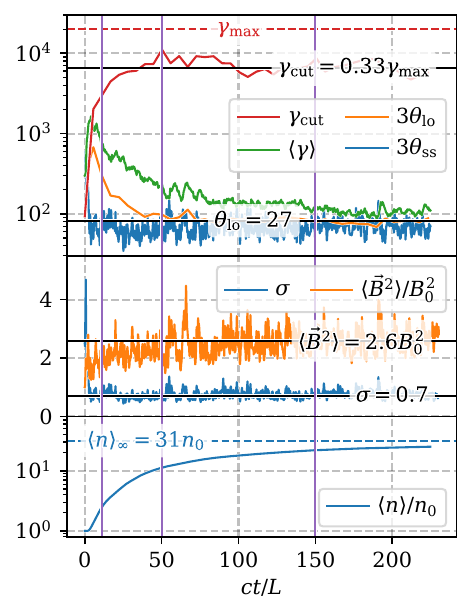}
    \caption{Timeseries of PIC box-averaged plasma quantities. Top: Cutoff Lorentz factor,~$\gamma_{\rm cut}$, as measured via~(\ref{eq:gcut}); overall average Lorentz factor,~$\langle \gamma \rangle$, as defined in~(\ref{eq:avggam}); temperature,~$\theta_{\rm lo}$, of the low-energy thermal hump as calculated from~(\ref{eq:thetalodef}); and theoretical equilibrium temperature,~$\theta_{\rm ss}$, defined by~(\ref{eq:thss}) with~$\eta$ set to~$1$. Middle: Instantaneous magnetization,~$\sigma$, and box-averaged magnetic energy density,~$\langle U_{\rm B} \rangle / U_{\rm B_0} = \langle \boldsymbol{B}^2 \rangle / B_0^2$. Bottom: Box-averaged particle density~$\langle n \rangle$. Vertical lines indicate the times pictured in Fig.~\ref{fig:pspecs}. Black horizontal lines denote time-averages taken over the second half~($ct/L > 120$) of the evolution. Time averages are used to evaluate the asymptotic pair yield~$\langle n \rangle_{\infty} = \langle \boldsymbol{B}^2 \rangle / (16 \pi \theta_{\rm lo} \sigma m_e c^2)$, that would be reached if the simulation could be integrated indefinitely.
    }
    \label{fig:lfs_and_sigh}
\end{figure}

We present timeseries of~$\gamma_{\rm cut}$,~$\theta_{\rm lo}$, and several other box-averaged plasma quantities in Fig.~\ref{fig:lfs_and_sigh}. In the figure, the instantaneous plasma magnetization is defined as~$\sigma = 3 \langle \boldsymbol{B}^2 \rangle / (16 \pi \langle \gamma \rangle \langle n \rangle m_e c^2)$. The rest of the plotted quantities are described in the caption.

Due to the onset of turbulent driving and magnetic field fluctuations~$\langle \delta \boldsymbol{B}^2 \rangle \simeq B_0^2$, the magnetization,~$\sigma$, initially jumps by roughly a factor of two from~$2.5$ to approximately~$5$. This does not last long, though, and is followed virtually immediately by a rapid drop, within the first several~$L/c$, in~$\sigma$ to its long-term steady-state value of roughly~$0.7$. This drop is mediated by an initial burst of particle acceleration and, hence, growth in~$\langle \gamma \rangle$.

The rapid equilibration of~$\sigma$ to its final steady-state value sets the theoretical temperature~$\theta_{\rm ss}$, through~(\ref{eq:thss}), to which the plasma eventually settles. Indeed, from Fig.~\ref{fig:lfs_and_sigh}, we see that the measured late-time temperature,~$\theta_{\rm lo} \simeq 27$, of the thermal component of the particle energy distribution matches the theoretical steady-state temperature,~$\theta_{\rm ss}$. Although~$\theta_{\rm ss}$ is set nearly immediately via~$\sigma$, the approach to the final thermal equilibrium unfolds more slowly, with,~$\theta_{\rm lo}$ and~$\langle \gamma \rangle$ taking much longer to relax to their respective late-time values of~$\theta_{\rm ss}$ and~$3 \theta_{\rm ss}$.

\begin{figure}
    \centering
    \includegraphics[width=\columnwidth]{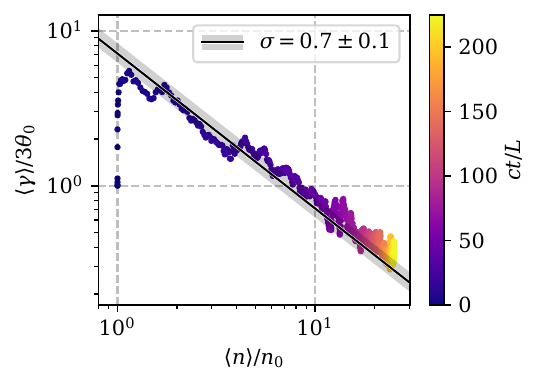}
    \caption{The mean Lorentz factor $\langle\gamma\rangle$ and plasma number density $\langle n \rangle$ evolve inversely proportionally, maintaining~$\sigma \simeq 0.7$ in the PIC simulation. The~$\sigma \simeq 0.7$ envelope corresponds to~$\langle \gamma \rangle / (3 \theta_0) = \tilde{\sigma}_0 n_0 / (\sigma \langle n \rangle)$, which is valid for constant~$\langle \boldsymbol{B}^2 \rangle$. We take~$\tilde{\sigma_0} = 2 \sigma_0 = 5$ to account for the fact that, once the magnetic fluctuations kick in,~$\langle \boldsymbol{B}^2 \rangle = \langle \delta \boldsymbol{B}^2 \rangle + B_0^2 \simeq 2 B_0^2$.
    }
    \label{fig:n_vs_avggam}
\end{figure}

After the initial few lightcrossing times,~$\sigma$ and~$\langle \boldsymbol{B}^2 \rangle$ remain roughly constant. This implies that~$\langle \gamma \rangle \propto \langle n \rangle^{-1}$, a dependence which we verify in Fig.~\ref{fig:n_vs_avggam}. Thus, following the initial and very rapid equilibration of~$\sigma$, the comparatively slow relaxation to the final state is characterized by an exchange of roles between~$\langle \gamma \rangle$ and~$\langle n \rangle$. Initially nonthermal particle acceleration dictates~$\sigma$ through a rapid increase in~$\langle \gamma \rangle$, but eventually pairs build up and the plasma cools such that, in the end, the number density~$\langle n \rangle$ maintains~$\sigma$ at its final equilibrium value. 

\section{Parameter exploration using 1D Fokker-Planck modeling}
\label{sec:fp}
The PIC simulations of Section~\ref{sec:pic} validate the expected sequence of events argued for on theoretical grounds in Section~\ref{sec:knexpectations}. However, because PIC simulations are expensive, they do not allow us to thoroughly explore the parameter space of IC- and pair-production-coupled radiative turbulence. Therefore, in this section, we present simplified (and, hence, cheaper) numerical models that we exploit to probe a broader parameter range. This allows us to constrain how the initial plasma state governs the final plasma temperature and magnetization.

In these simplified models, we discard self-consistent particle acceleration by the electromagnetic fields (and, with it, the need to run expensive PIC simulations). We assume instead that particle acceleration obeys a Fokker-Planck (FP) equation of the form (cf.\ equation~\ref{eq:fp})
\begin{align}
    \partial_t f &= \partial_\gamma \left( D \partial_\gamma f \right) - \partial_\gamma \left[ \left( 2 D / \gamma + A \right) f \right] \, \notag \\
    &+ \mathrm{IC\, cooling} + \mathrm{pair\, production} \, ,
    \label{eq:fpfull}
\end{align}
where~$D(\gamma) = (\gamma^2 / t_{\rm acc}) \exp(-\gamma/\gamma_{\rm max})$,~$A(\gamma) = 0$, and the exponential cutoff at~$\gamma_{\rm max}$ in~$D(\gamma)$ mimics a system-size cutoff Lorentz factor. We note that, here, unlike in Section~\ref{sec:knexpectations}, we have not chosen to model IC cooling as an advective term in~(\ref{eq:fpfull}): that is,~$A(\gamma) = 0 \neq A_{\rm IC}(\gamma)$. Such a treatment would imply that cooling is always continuous, whereas, for~$\gamma \gg \gamma_{\rm KN}$, it is discrete. We employ instead a completely rigorous treatment of pair production and IC cooling, identical to that used in our PIC simulations \citep{mehlhaff_etal_2024}.

To model feedback from the plasma state on particle acceleration, we update~$t_{\rm acc}$ in time according to equation~(\ref{eq:taccconstraint}) based on the instantaneous plasma magnetization~$\sigma$. Since~$\sigma$ involves the magnetic field, which is not tracked by our FP models, we need a suitable prescription for setting it. We take inspiration here from our PIC simulations (Fig.~\ref{fig:n_vs_avggam}), in which the magnetic field remains roughly constant and, hence,~$\sigma$ virtually only changes in response to the average particle energy,~$\langle \gamma \rangle$, and number growth,~$\langle n \rangle / n_0$. We therefore adopt, in our FP models,~$\sigma = 3 \theta_{0} n_0 \sigma_0 / (\langle \gamma \rangle \langle n \rangle)$.

\begin{figure}
    \centering
    \includegraphics[width=\columnwidth]{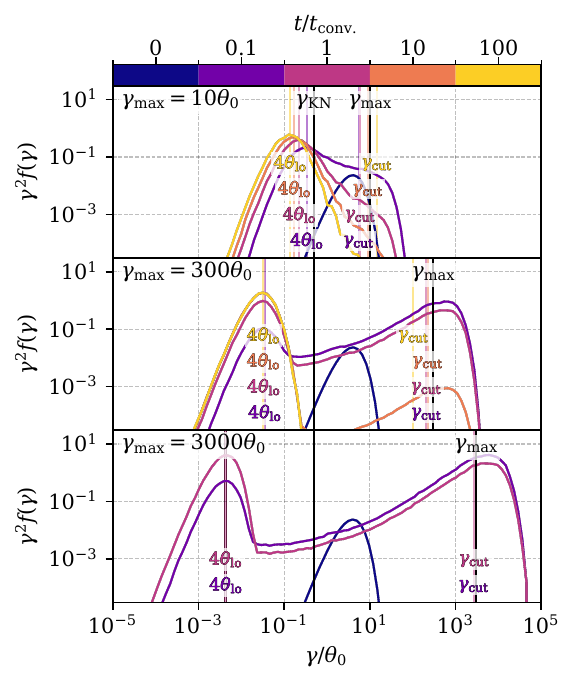}
    \caption{Particle energy distribution snapshots from our FP runs with~$\gamma_{\rm max} / \theta_{0} = 10$,~$300$, and~$3000$. The convergence time,~$t_{\rm conv}$, is measured as described in Section~\ref{sec:fp}. Instantaneous~$\theta_{\rm lo}$ and~$\gamma_{\rm cut}$ values are measured using~(\ref{eq:gcut}) and~(\ref{eq:thetalodef}). The run with~$\gamma_{\rm max} = 3000 \theta_0$ had an extremely long~$t_{\rm conv}$ (Fig.~\ref{fig:eq_tconverge}), making it too expensive to evolve through~$t/t_{\rm conv} = 10$ and~$100$.}
    \label{fig:fp}
\end{figure}

We start our FP runs from a Maxwell-Jüttner particle energy distribution with ultrarelativistic initial temperature~$\theta_{\rm 0} \gg 1$. The exact value of~$\theta_{\rm 0}$ is arbitrary, since, in the ultrarelativistic limit, it merely supplies a fiducial energy scale. We choose a value that is high enough to ensure that all particles remain ultrarelativistic, even when cooled substantially (e.g., lower panel of Fig.~\ref{fig:fp}).

To set up the FP models, we must further specify the three constants,\footnote{We omit~$U_{\rm rad}$ and~$\epsilon_{\rm rad}$ here, since they are set, respectively, by~$\gamma_{\rm cool}$ and~$\gamma_{\rm KN}$.}~$\gamma_{\rm cool}$,~$\gamma_{\rm KN}$, and~$\tau_{\gamma\gamma}$, plus initial values,~$\theta_{\rm ss,0}$,~$L/(ct_{\rm acc,0})$, and~$\sigma_0$, of the three time-dependent quantities,~$\theta_{\rm ss}$,~$L/(ct_{\rm acc})$, and~$\sigma$. Here,~$\theta_{\rm ss}$ and~$L/(ct_{\rm acc})$ inherit their time-dependence through~$\sigma$, per~(\ref{eq:thss}) and~(\ref{eq:taccconstraint}). These six parameters can be related by noting, from~(\ref{eq:thss}) and~(\ref{eq:taccconstraint}), that~$\theta_{\rm ss} / \gamma_{\rm cool} = L / (c t_{\rm acc})$, which implies that
\begin{align}
    \tau_{\gamma\gamma} = \frac{3}{5} \frac{\gamma_{\rm KN}}{\gamma_{\rm cool}} = \frac{3}{5} \frac{\gamma_{\rm KN}}{\theta_{\rm ss}} \frac{L}{c t_{\rm acc}} \, .
    \label{eq:choose3}
\end{align}
Though this identity is valid at all times, we use it mainly to set the initial FP parameters. Equation~(\ref{eq:choose3}) implies that choosing any three of~$\tau_{\gamma\gamma}$,~$\gamma_{\rm KN}$,~$\theta_{\rm ss,0}$, and~$L/(c t_{\rm acc,0})$ constrains the fourth, and, by extension,~$\gamma_{\rm cool}$ and~$\sigma_0$.

In order for QED effects to be relevant,~$\theta_{\rm ss,0}$ must exceed (or at least not be much smaller than) $\gamma_{\rm KN}$; otherwise, the system stays trapped in the Thomson radiative regime. This implies, through~(\ref{eq:choose3}), ignoring factors of order unity, that~$L / (c \tau_{\gamma \gamma}) > t_{\rm acc,0}$. The left-hand side of this inequality,~$L / (c \tau_{\gamma \gamma})$, is the main QED-related timescale: it is both the typical lifetime of above-threshold photons and the fastest-possible IC cooling time, achieved at particle energies~$\gamma \sim \gamma_{\rm KN}$ (Section~\ref{sec:primer}). Equation~(\ref{eq:choose3}) therefore tells us that the initial acceleration timescale must always be faster than that associated with radiative cooling in order to push the system into the pair-producing regime. 

In our FP runs, we choose~$L/(ct_{\rm acc,0}) = 10$. The above remarks then constrain us to choose~$\tau_{\rm \gamma \gamma} < L / (c t_{\rm acc,0}) = 10$. We have simulated~$\tau_{\gamma \gamma} = 1$,~$3$, and~$10$, but we present here only the runs for which~$\tau_{\gamma\gamma} = 3$. For simplicity, we also demand that~$\theta_{\rm ss,0}$ be equal to~$\theta_{\rm 0}$. This then sets~$\gamma_{\rm KN}$ and~$\gamma_{\rm cool}$ through~(\ref{eq:choose3}) as well as~$\sigma_0$ through~(\ref{eq:taccconstraint}) with~$\eta = 1$. Specifically,~$\gamma_{\rm KN} = 0.5 \theta_{\rm ss,0}$,~$\gamma_{\rm cool} = 0.1 \theta_{\rm ss,0}$, and~$\sigma_0 = 60$.

By choosing~$\theta_{\rm 0} = \theta_{\rm ss,0} > \gamma_{\rm KN}$, many particles start out above pair-production threshold, which ignites pair production directly from our initial conditions. We are willing to accept this because avoiding it by choosing~$\theta_{\rm 0} < \theta_{\rm ss,0}$ would obfuscate the meaning of~$\theta_{\rm ss,0}$. As particle acceleration proceeded,~$\sigma$ would drop in time, rendering the effective~$\theta_{\rm ss}$, by the time pair production kicked in, different from~$\theta_{\rm ss,0}$.

The last parameter we need to specify is~$\gamma_{\rm max}$. This is the main parameter that we scan throughout our FP campaign, pushing it much higher (with respect to~$\theta_{\rm 0}$) than is possible in PIC simulation. We conduct runs with~$\gamma_{\rm max} \in \left\{ 3, 10, 30, 100, 300, 1000, 3000 \right\} \theta_{\rm 0}$.

The QED part of the time evolution of our FP simulations is identical to that described by \citet{mehlhaff_etal_2024}. The diffusive part of the update is, however, new to this work. We provide a technical description of this novelty in Appendix~\ref{sec:fptech}.

We illustrate the evolution of our FP models in Fig.~\ref{fig:fp}. They go through the same sequence of states (Section~\ref{sec:knexpectations}) as our PIC simulation. First, particles are rapidly accelerated to Lorentz factors~$\gamma>\gamma_{\rm KN}$ where they emit pair-producing gamma rays. Next, pairs gradually accumulate, loading the plasma magnetization~$\sigma$ and, with it, dampening the efficiency of further particle acceleration. Finally, the plasma settles into a thermal distribution at temperature~$\theta_{\rm ss,f}$. 

Two important outputs of our FP models are the final temperature,~$\theta_{\rm ss,f}$, and the time,~$t_{\rm conv}$, it takes the system to converge to the long-term equilibrium. We measure these two quantities as follows. First, we define~$t_{\rm conv}$ as the time when the overall average Lorentz factor,~$\langle \gamma \rangle$, comes within a factor of two of the instantaneous steady-state temperature,~$\theta_{\rm ss}$: that is,~$\langle \gamma \rangle(t_{\rm conv}) \equiv 2 \times 3\theta_{\rm ss}(t_{\rm conv})$. This roughly corresponds (Fig.~\ref{fig:fp}) to the moment when half the plasma energy is carried by the low-energy thermalized particles. We define~$\theta_{\rm ss,f}$ as the value of~$\theta_{\rm lo}$ time-averaged over the second (converged) half of each FP run. The main goal of the following subsections is to show how the FP parameters dictate~$\theta_{\rm ss,f}$ and~$t_{\rm conv}$.

\subsection{Analysis of the final steady-state temperature}
\label{sec:thetalo}
At the beginning of each FP simulation, particle acceleration occurs on timescales,~$t_{\rm acc,0}$, much faster than the most rapid radiative cooling timescale,~$L/(c\tau_{\gamma\gamma})$. We therefore expect the tail of the particle energy distribution to initially run away, as predicted by~(\ref{eq:fpsolnkn}). The cutoff in the particle energy distribution must then be regulated, at least at first, by a mechanism other than radiative cooling. The only mechanism available in our FP models is the imposed cutoff in the diffusion coefficient at Lorentz factor~$\gamma_{\rm max}$. We therefore anticipate that the final plasma temperature,~$\theta_{\rm ss,f}$, be sensitive to~$\gamma_{\rm max}$. 

\begin{figure}
    \centering
    \includegraphics[width=\columnwidth]{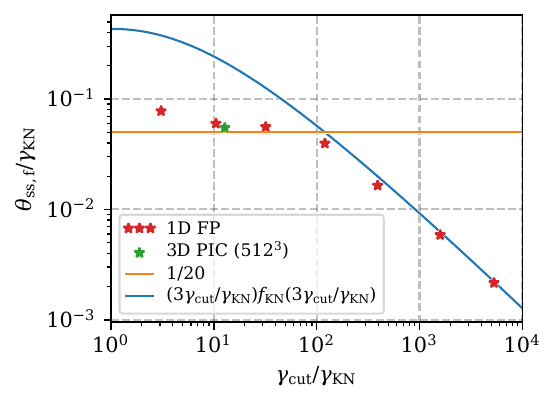}
    \caption{Dependence of the final plasma temperature,~$\theta_{\rm ss,f}$, on~$\gamma_{\rm cut}$, the main parameter explored in our FP campaign. The formulae from~(\ref{eq:thssformula}) and the measurement from our PIC simulation are shown for reference.}
    \label{fig:eq_thetalo}
\end{figure}

This is indeed the case. In Fig.~\ref{fig:eq_thetalo}, we show how~$\theta_{\rm ss,f}$ varies as a function of~$\gamma_{\rm cut} / \gamma_{\rm KN}$.
To compare with PIC results, we plot~$\theta_{\rm ss,f}$ in terms of the measured cutoff Lorentz factor~$\gamma_{\rm cut}$, defined by~(\ref{eq:gcut}), instead of in terms of the formal parameter~$\gamma_{\rm max}$. These two are, however, closely related, differing only by order-unity factors from each other. 
The data points in Fig.~\ref{fig:eq_thetalo} follow the trend:
\begin{align}
    \frac{\theta_{\rm ss,f}}{\gamma_{\rm KN}} = \begin{cases}
        \frac{1}{20}, & \gamma_{\rm cut} \lesssim 100 \gamma_{\rm KN} \\
        \left( \frac{3\gamma_{\rm cut}}{\gamma_{\rm KN}} \right) f_{\rm KN}\left( \frac{3\gamma_{\rm cut}}{\gamma_{\rm KN}} \right), & \rm otherwise
    \end{cases} \, .
    \label{eq:thssformula}
\end{align}
These scalings can be phenomenologically explained as follows.

The regime~$\gamma_{\rm cut} \gg 100 \gamma_{\rm KN}$ emerges from the balance between the radiative cooling timescale,~$t_{\rm cool}(\gamma)$ (independent of~$\sigma$; equation~\ref{eq:tcool}), and the diffusive acceleration timescale,~$t_{\rm acc}(\sigma)$, dictated by~$\sigma$ per~(\ref{eq:taccconstraint}). Before substantial pair production can occur,~$\sigma$ is regulated in our FP runs, just like in our PIC simulation, through particle acceleration. Initially,~$\langle \gamma \rangle$ increases, lowering~$\sigma = 3 \theta_0 n_0 \sigma_0 / (\langle n \rangle \langle \gamma \rangle)$. We empirically observe that, in the~$\gamma_{\rm cut} \gg 100 \gamma_{\rm KN}$ regime, the FP models all regulate~$\sigma$ in this pre-pair-production stage so as to enforce the equality~$t_{\rm acc}(\sigma) = t_{\rm cool}(\xi \gamma_{\rm cut})$. Here,~$\xi$ is the constant order-unity factor by which the precise value of~$\gamma$ where this equality is satisfied differs from the empirically measured~$\gamma_{\rm cut}$. We find~$\xi \simeq 3$.

As noted in Section~\ref{sec:primer}, the cooling time is non-monotonic in~$\gamma$:~$d t_{\rm cool} / d \gamma > 0$ when~$\gamma \gg \gamma_{\rm KN}$ and~$d t_{\rm cool} / d \gamma < 0$ when~$\gamma \ll \gamma_{\rm KN}$. Thus, by self-regulating such that~$t_{\rm acc}(\sigma) = t_{\rm cool}(3 \gamma_{\rm cut})$, the system also defines a second critical Lorentz factor, paired with~$\gamma_{\rm cut}$ -- but, unlike~$\gamma_{\rm cut}$, less than~$\gamma_{\rm KN}$ -- for which the equality~$t_{\rm acc} = t_{\rm cool}$ is also satisfied. This second Lorentz factor is precisely~$\theta_{\rm ss,f}$. We can thus phenomenologically predict~$\theta_{\rm ss,f}$ by equating~$t_{\rm cool}(\theta_{\rm ss,f}) = t_{\rm acc}(\sigma) = t_{\rm cool}(3 \gamma_{\rm cut})$. Noting that~$\theta_{\rm ss,f} \ll \gamma_{\rm KN}$ transforms the left-hand-side of this equality into~$ct_{\rm cool}(\theta_{\rm ss,f})/L \simeq \gamma_{\rm cool} / \theta_{\rm ss,f}$. Meanwhile, the right-hand-side, by~(\ref{eq:tcool}), is just~$ct_{\rm cool}(3 \gamma_{\rm cut})/L = \gamma_{\rm cool} / [3 \gamma_{\rm cut} f_{\rm KN}(3 \gamma_{\rm cut} / \gamma_{\rm KN})]$. Rearranging, we see that,~$\theta_{\rm ss,f} / \gamma_{\rm KN} = (3 \gamma_{\rm cut} / \gamma_{\rm KN}) f_{\rm KN}(3 \gamma_{\rm cut} / \gamma_{\rm KN})$.

We can then understand the regime~$\gamma_{\rm cut} \lesssim 100 \gamma_{\rm KN}$ by noting that the form for~$\theta_{\rm ss,f}$ obtained above cannot continue indefinitely to smaller~$\gamma_{\rm cut} / \gamma_{\rm KN}$. Doing so would eventually predict~$\theta_{\rm ss,f} / \gamma_{\rm KN} \sim 1$. However, this is forbidden because having~$\theta_{\rm ss,f} \sim \gamma_{\rm KN}$ would lead to continued pair production, invalidating the final steady state. The final temperature must be sufficiently below~$\gamma_{\rm KN}$ to render pair production virtually nonexistent. Evidently, the system requires~$\theta_{\rm ss,f}$ to be no larger than roughly~$\gamma_{\rm KN}/20$. We note that an exponentially small amount of pair production still persists after~$t=t_{\rm conv}$ in our FP runs, slowly pushing the thermal particles to still colder temperatures. However, this occurs on timescales that are already much longer than~$t_{\rm conv}$ and that become even slower as the cooling continues.

\subsection{Analysis of the convergence time}
\label{sec:tconv} 
In our FP runs, as in our PIC simulation, the final magnetization,~$\sigma_{\rm f}$, is decided very early. Initially, vigorous particle acceleration leads to a rapid rise in~$\langle \gamma \rangle$, setting~$\sigma_{\rm f}$ within the first several~$L/c$. In the subsequent, much slower approach to the final state,~$\sigma$ remains roughly constant. As pairs build up,~$\langle \gamma \rangle$ and~$\langle n \rangle$ merely exchange roles, obeying~$\langle \gamma \rangle \propto 1/\langle n \rangle$, and, hence, preserving~$\sigma$.

In this sequence of events, the slow accumulation of pairs -- which dominates the time to reach the final state -- is governed by the radiative cooling time,~$t_{\rm cool}$, as well as the lifetime of above-threshold photons,~$\sim L/(c\tau_{\gamma\gamma})$. Both of these timescales are proportional to~$L/(c\tau_{\gamma\gamma})$ times a dimensionless function of (particle or photon) energy. Hence, we should expect the time,~$t_{\rm conv}$, to converge to the final steady state to be proportional to~$L/(c\tau_{\gamma\gamma})$.

\begin{figure}
    \centering
    \includegraphics[width=\columnwidth]{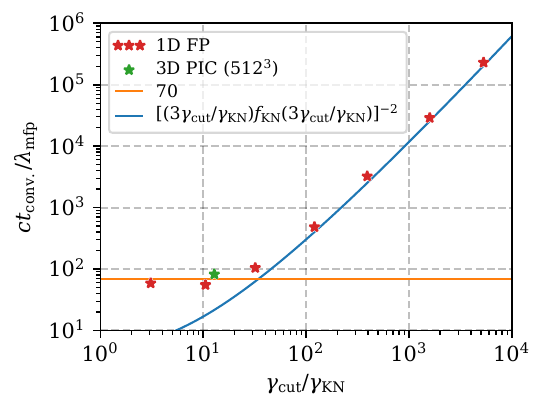}
    \caption{Dependence of the convergence time,~$t_{\rm conv}$, on~$\gamma_{\rm cut}$, the main parameter explored in our FP campaign. The formulae from~(\ref{eq:tconvformula}) and the convergence time of our PIC simulation are shown for reference.}
    \label{fig:eq_tconverge}
\end{figure}

These remarks are why, in Fig.~\ref{fig:eq_tconverge}, we normalize the measured~$t_{\rm conv}$ from each of our simulations to~$\lambda_{\rm mfp}/c \equiv L / (c \tau_{\gamma\gamma})$. For each value of~$\gamma_{\rm cut}/\gamma_{\rm KN}$, changing~$\lambda_{\rm mfp} = L / \tau_{\gamma\gamma}$ leads to a different absolute convergence time (e.g., different~$ct_{\rm conv}/L$) but leaves~$ct_{\rm conv}/\lambda_{\rm mfp}$ unchanged. We have verified this explicitly using our runs (not presented) with different~$\tau_{\gamma\gamma}$.

The measured~$t_{\rm conv}$ values of Fig.~\ref{fig:eq_tconverge}, like those of~$\theta_{\rm ss,f}$ in Fig.~\ref{fig:eq_thetalo}, follow two separate trends depending on~$\gamma_{\rm cut} / \gamma_{\rm KN}$. We find that
\begin{align}
    \frac{ct_{\rm conv}}{\lambda_{\rm mfp}} = \begin{cases}
        70, & \gamma_{\rm cut} \lesssim 100 \gamma_{\rm KN} \\
        \left[ \left( \frac{3\gamma_{\rm cut}}{\gamma_{\rm KN}} \right) f_{\rm KN}\left( \frac{3\gamma_{\rm cut}}{\gamma_{\rm KN}} \right) \right]^{-2}, & \rm otherwise
    \end{cases} \, .
    \label{eq:tconvformula}
\end{align}
We next supply phenomenological arguments to explain each of these scalings.

First, the case~$\gamma_{\rm cut} \gg 100 \gamma_{\rm KN}$ is very extreme. There, the early phases of the FP runs show that a good fraction of the initial particles are rapidly (within a few~$L/c$) accelerated up to~$\gamma \simeq \xi \gamma_{\rm cut} \simeq 3 \gamma_{\rm cut}$. Subsequently, radiative cooling removes energy from these particles at the rate~$\dot{\varepsilon}_{\rm depart} \simeq 3 \gamma_{\rm cut} n_0 m_e c^2 / t_{\rm cool}(3 \gamma_{\rm cut})$. At such high Lorentz factors, virtually all Comptonized photons lie above pair-production threshold, are absorbed, and, thus, do not remove energy from the system. Energy is therefore losslessly reprocessed from particles with~$\gamma \sim 3 \gamma_{\rm cut}$ down to (a larger number of) particles with~$\gamma \sim \gamma_{\rm KN}$. Then, from~$\gamma \sim \gamma_{\rm KN}$ down to~$\gamma \sim \theta_{\rm ss,f} \ll \gamma_{\rm KN}$, the IC-scattered photons fall below pair-production threshold, and -- no longer subject to gamma-ray absorption -- carry energy out of the plasma. Thus, while energy is removed from the high-energy particles at the rate~$\dot{\varepsilon}_{\rm depart}$, it arrives at~$\theta_{\rm ss,f}$ at a lower rate, penalized by the losses incurred from~$\gamma_{\rm KN}$ to~$\theta_{\rm ss,f}$, of~$\dot{\varepsilon}_{\rm arrive} \sim (\theta_{\rm ss,f} / \gamma_{\rm KN}) \dot{\varepsilon}_{\rm depart} \sim (\theta_{\rm ss,f} / \gamma_{\rm KN}) [3 \gamma_{\rm cut} n_0 m_e c^2/ t_{\rm cool}(3 \gamma_{\rm cut})]$. The convergence time is reached when energy of roughly~$3 \gamma_{\rm cut} n_0 m_e c^2$ accrues at~$\gamma \sim \theta_{\rm ss,f}$. Thus,~$t_{\rm conv} \sim 3 \gamma_{\rm cut} n_0 m_e c^2 / \dot{\varepsilon}_{\rm arrive} \sim (\gamma_{\rm KN} / \theta_{\rm ss,f}) t_{\rm cool}(3 \gamma_{\rm cut})$. Plugging in~(\ref{eq:thssformula}) and~(\ref{eq:tcool}) and ignoring order-unity factors yields,~$c t_{\rm conv} / \lambda_{\rm mfp} \sim [(3 \gamma_{\rm cut} / \gamma_{\rm KN}) f_{\rm KN}(3 \gamma_{\rm cut} / \gamma_{\rm KN})]^{-2}$.

Conversely, in the limit that~$\gamma_{\rm cut} \ll 100 \gamma_{\rm KN}$, the initial burst of particle acceleration in the FP models is much less efficient (e.g., upper panel of Fig.~\ref{fig:fp}). Most of the energy remains in the initial thermal hump, which just cools down from its initial temperature,~$\theta_{0}$, to the final one,~$\theta_{\rm ss,f} \sim \gamma_{\rm KN}$. The time for this cooling to occur is just~$t_{\rm cool}(\gamma_{\rm KN}/20) \simeq 30 \lambda_{\rm mfp} / c$. This explains why~$t_{\rm conv}$ should asymptote to a~$\gamma_{\rm cut}$-independent value in the~$\gamma_{\rm cut} \ll 100 \gamma_{\rm KN}$ limit, but ignores factors of order unity. For quantitative estimates, we use the value~$c t_{\rm conv} / \lambda_{\rm mfp} \simeq 70$, which is in better agreement with the data of Fig.~\ref{fig:eq_tconverge}.

\subsection{Discussion of the limit~$\gamma_{\rm cut} \lesssim 100 \gamma_{\rm KN}$}
\label{sec:gcutlow}
In Section~\ref{sec:astroapps}, we consider the implications of Klein-Nishina turbulence on the pair content of FSRQ jets. There, we identify~$\gamma_{\rm cut} \lesssim 100 \gamma_{\rm KN}$ as the relevant regime in such jets. It is therefore worth exploring the consequences of this regime on the final plasma parameters in more depth.

In this limit,~$\theta_{\rm ss,f} \sim \gamma_{\rm KN} / 20$ (equation~\ref{eq:thssformula}). Plugging this temperature into~(\ref{eq:thss}), we estimate the final plasma magnetization,~$\sigma_{\rm f}$, as
\begin{align}
    \frac{3 \gamma_{\rm KN}}{10 \gamma_{\rm cool}} = \frac{\tau_{\gamma\gamma}}{2} = \sqrt{\frac{\sigma_{\rm f}^3}{1+\sigma_{\rm f}}} \, .
    \label{eq:sigtaugg}
\end{align}
This is an implicit expression for the final plasma magnetization,~$\sigma_{\rm f}$, in terms of the optical depth,~$\tau_{\gamma\gamma}$. In the optically thick regime,~$\tau_{\gamma\gamma} \gg 1$, it reduces to the very simple result,~$\sigma_{\rm f} \simeq \tau_{\gamma\gamma} /2$. Equation~(\ref{eq:sigtaugg}) makes no reference to the initial plasma parameters (e.g.,~$\sigma_0$ and~$\theta_0$), implying that the final plasma state is entirely independent of the initial one. Remarkably,~$\sigma_{\rm f}$ in~(\ref{eq:sigtaugg}) does not even depend on plasma parameters at all; it is determined exclusively by properties of the seed radiation field through~$\tau_{\gamma\gamma}$.

Furthermore, when~$\gamma_{\rm cut} \lesssim 100 \gamma_{\rm KN}$, equation~(\ref{eq:tconvformula}) implies that the convergence time is roughly~$t_{\rm conv} \simeq 70 \lambda_{\rm mfp}/c = 70 L / (c \tau_{\gamma\gamma})$. Thus, higher~$\tau_{\gamma\gamma}$ corresponds to a faster approach to equilibrium (with respect to~$L/c$), but to a less pair-loaded plasma, giving, according to~(\ref{eq:sigtaugg}), higher~$\sigma_{\rm f}$.

\section{Turbulence-powered pair enrichment of FSRQ jets}
\begin{figure*}
    \centering
    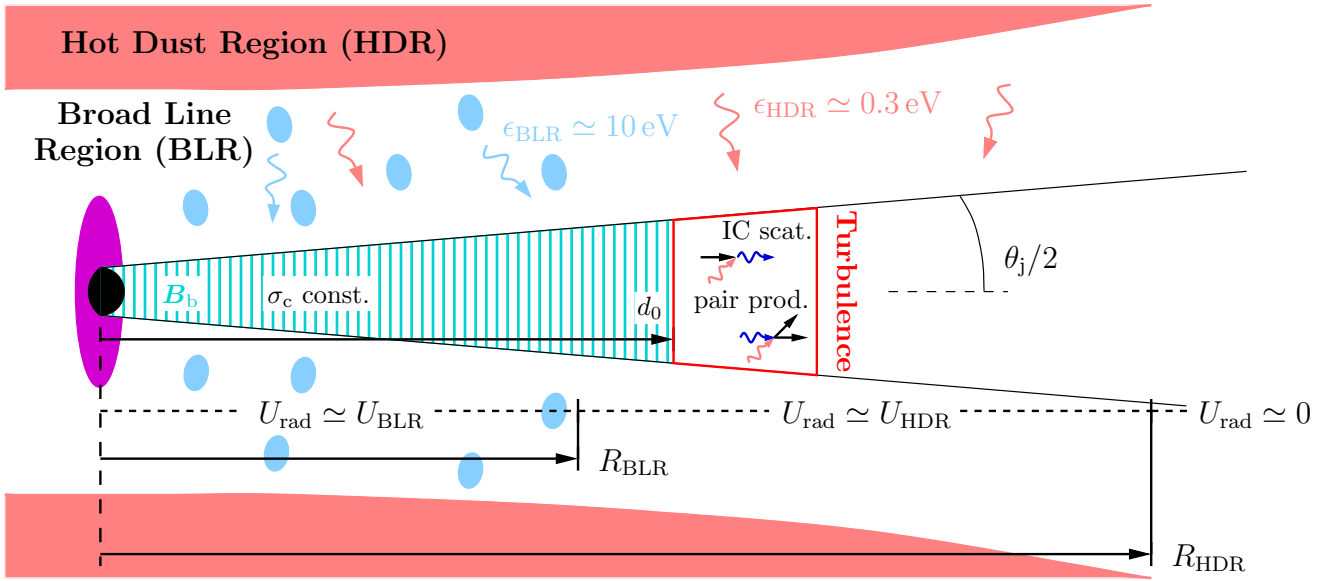
    \caption{Cartoon of our global model of turbulence embedded in an FSRQ jet \citep[cf.][]{sikora_etal_1994,urry_padovani_1995, ghisellini_tavecchio_2009}. Turbulence couples exclusively to the photons from the HDR (BLR) for~$R_{\rm BLR}<\jetdist_0<R_{\rm HDR}$~($\jetdist_0 < R_{\rm BLR}$). The case of HDR coupling is shown. Symbol definitions are given in the text.}
    \label{fig:jet}
\end{figure*}

The main goal of this section is to determine whether turbulence, coupled to IC radiation and pair production as studied in this work, might be a significant source of electron-positron pairs in jets of flat-spectrum radio quasars (FSRQs). To do so, we introduce a global model, illustrated in Fig.~\ref{fig:jet}, wherein a patch of turbulent plasma travels outward in the jet at the local relativistic jet velocity,~$v_{\rm j} = (1 - 1/\Gamma_{\rm j}^2)^{1/2} c \simeq c$. The jet is assumed conical with full opening angle~$\theta_{\rm j} = 1/\jetopen \Gamma_{\rm j}$, where~$\jetopen \simeq5$ is an observationally determined constant \citep[][]{pushkarev_etal_2009}. Throughout this section, unprimed quantities refer to the rest frame of the central engine and primed quantities to the comoving frame of turbulent jet plasma. Though needed to transform between these frames, the bulk jet Lorentz factor,~$\Gamma_{\rm j}$, cancels in our final results. 
\label{sec:astroapps}

We suppose that turbulence is triggered (by, e.g., a macroscopic instability) in the jet-comoving frame at a distance~$\jetdist=\jetdist_0$ from the central engine, with outer scale,~$L$, of order the jet width:~$L\sim \theta_{\rm j}\jetdist_0 \sim \jetdist_0/\jetopen \Gamma_{\rm j}$. To compare with our preceding results on pair-plasma turbulence, we assume the jet is comprised only of electrons and positrons. We discuss rough expectations for the case of an electron-ion jet in Section~\ref{sec:fsrqsummary}. We also assume that turbulence does not coincide with strong bulk acceleration, so that~$\Gamma_{\rm j}$ is constant. In reality, though, the Compton rocket effect \citep{odell_1981, sikora_etal_1996, vuillaume_etal_2015} could impart a strong net momentum to the plasma; turbulent dissipation could also create a large-scale (magnetic) pressure gradient along the jet axis, leading to bulk acceleration \citep{drenkhahn_2002, giannios_uzdensky_2019}. We relegate a proper analysis of the link between turbulent dissipation and bulk acceleration to a future work, focusing here just on that between the dissipation and pair content of the jet.

Once triggered, turbulence couples to the BLR or HDR background radiation via IC scattering and pair production. \citet{mehlhaff_etal_2021} summarize the quantitative aspects of the BLR and HDR soft-photon baths that are relevant here. The BLR and HDR shine radiation at respective characteristic photon energies~$\epsilon_{\rm BLR} = 10 \, \rm eV$ and~$\epsilon_{\rm HDR} = 0.3 \, \rm eV$. The energy densities,~$U_{\rm BLR}$ and~$U_{\rm HDR}$, of the resulting radiation bathing the jet are both roughly constant in~$\jetdist$ as long as~$\jetdist$ is smaller than the characteristic size,~$R_{\rm BLR}$ and~$R_{\rm HDR}$, of the respective region. Beyond~$R_{\rm BLR}$ or~$R_{\rm HDR}$, the corresponding seed-photon energy density falls off rapidly in~$\jetdist$. The BLR is smaller and more intense than the HDR, with~$0.1 \, \rm pc \sim R_{\rm BLR} < R_{\rm HDR} \sim 4 \, pc$ and~$U_{\rm BLR} > U_{\rm HDR}$ for~$\jetdist<R_{\rm BLR}$.

Given these orderings, we consider two separate possibilities: either turbulence is triggered at~$\jetdist_0 < R_{\rm BLR}$ and, hence, couples exclusively to the more intense BLR radiation, or turbulence is triggered at~$R_{\rm BLR} < \jetdist_0 < R_{\rm HDR}$ and couples exclusively to the HDR photons (which become dominant at~$\jetdist \gtrsim 1 \, \rm pc$). Mixed cases where, for example, turbulence is triggered at~$\jetdist_0 < R_{\rm BLR}$ but does not converge before being swept out of the BLR, may very well occur in nature, but we ignore them here to streamline our analysis. To summarize, we assume
\begin{align}
    U_{\rm rad} = \begin{cases}
        U_{\rm BLR} \, , & \jetdist_0 < R_{\rm BLR} \\
        U_{\rm HDR} \, , & R_{\rm BLR} < \jetdist_0 < R_{\rm HDR} \\
        0 \, , & \mathrm{otherwise}
    \end{cases}
    \label{eq:uradcases}
\end{align}
and
\begin{align}
    \epsilon_{\rm rad} = \begin{cases}
        \epsilon_{\rm BLR} \, , & \jetdist_0 < R_{\rm BLR} \\
        \epsilon_{\rm HDR} \, , & R_{\rm BLR} < \jetdist_0 < R_{\rm HDR} \\
        0 \, , & \mathrm{otherwise}
    \end{cases} \, .
    \label{eq:eradcases}
\end{align}
Figure~\ref{fig:jet} illustrates the case~$R_{\rm BLR} < z_0 < R_{\rm HDR}$.

Within our model, necessary conditions for turbulence-powered pair enrichment of the jet are:
\begin{enumerate}
\item That the convergence time,~$t_{\rm conv}$, be shorter than the time,~$t_{\rm turb}$, for turbulence to subside;
\label{en:tconvtend}
\item That~$t_{\rm conv}$ be shorter than the time,~$\tesc$, for the turbulent region to be ejected from the seed radiation field; and
\label{en:tconvtesc}
\item That the resulting pair yield, characterized as the ratio of the final-to-initial electron+positron number densities,~$n_{\rm f}/n_0$, exceed unity.
\label{en:yield}
\end{enumerate}
To facilitate transparent comparisons to our strictly local turbulence analysis (Sections~\ref{sec:pic} and~\ref{sec:fp}), we impose an additional \textit{locality condition}. We demand that~$t_{\rm conv}$ be shorter than the time for turbulence to travel far enough in~$\jetdist$ for the background plasma properties (e.g., density and magnetic field strength) to substantially change due to jet expansion. Namely, we require~$t_{\rm conv} < t_{\times 2} = 2 \jetdist_0 / c$, where~$t_{\times 2}$ is the time for turbulence to reach the location,~$\jetdist=2\jetdist_0$, where the jet is twice as wide as at~$\jetdist_0$.

We stress that the locality condition is only enforced to permit direct application of the quantitative findings of our local turbulence study. It may not actually be required for turbulence to substantially boost the jet pair content. Nevertheless, taking quantitative account of jet expansion would require additional study beyond our present scope. One could, for example, conduct PIC simulations similar to those of Section~\ref{sec:pic} in an expanding-box framework \citep{sironi_narayan_2015, tenerani_velli_2017, bott_etal_2021}.

The rest of this section is organized as follows. We address the ejection~($t_{\rm conv}<\tesc$) and locality~($t_{\rm conv}<t_{\times 2}$) conditions in~\ref{sec:tconvtesc}. Then, in~\ref{sec:yield}, we estimate the pair yield. Finally, we circle back to discuss whether turbulence converges before it subsides~($t_{\rm conv}<t_{\rm turb}$) as part of our discussion of this section in~\ref{sec:fsrqsummary}.

\subsection{Ejection and locality conditions}
\label{sec:tconvtesc}
To evaluate the ejection and locality conditions, we need to estimate the timescales~$\tesc$ and~$t_{\rm conv}$. We begin with~$\tesc$. For illumination by the BLR and HDR, respectively, we have~$\tesc_{,\rm BLR} = (R_{\rm BLR}-\jetdist_0)/c$ and~$\tesc_{,\rm HDR} = (R_{\rm HDR}-\jetdist_0)/c$.

The ejection time,~$\tesc_{,i}$ (where~$i =$ BLR or HDR), becomes formally equal to~$t_{\times 2}$ when~$\jetdist_0 = R_i / 3$. For~$\jetdist_0$ closer in than this,~$\tesc_{,i} > t_{\times 2}$, while for~$\jetdist_0$ farther away,~$\tesc_{,i} < t_{\times 2}$. To allow turbulence a maximum of time to bathe in the ambient radiation, we restrict our assumption on~$\jetdist_0$, requiring it to be less than or comparable to~$R_i/3$. That way, as long as the locality condition,~$t_{\rm conv} < t_{\times 2}$, is satisfied, the ejection condition,~$t_{\rm conv} < \tesc_{,i}$, is automatically respected.

We next move on to estimate~$t_{\rm conv}$, which, according to the results of Section~\ref{sec:tconv}, follows from the particle acceleration cutoff Lorentz factor,~$\gamma_{\rm cut}$, normalized by~$\gamma_{\rm KN}$. In FSRQs, the IC gamma-ray brightness does not generally exceed the lower-energy synchrotron luminosity by more than a factor of~$100$ or so. If both spectra are produced by the same particle population -- and barring strong beaming or Klein-Nishina effects, the latter of which occurs at energies beyond the spectral peak -- the energy density,~$U_{\rm i}'$, of background radiation in the emission zone cannot exceed the magnetic field energy density,~$U_{\rm B}'$, by more than the same factor of~$100$ \citep{sikora_etal_2009}.

This fact can be translated into a constraint on~$\gamma_{\rm cut}$ as follows. Synchrotron radiation is quantitatively similar to Thomson IC cooling. For an isotropic particle pitch-angle distribution, the per-particle radiated synchrotron power obeys~(\ref{eq:eradthom}) but with the seed-photon energy density replaced by the magnetic energy density. Hence, there is a synchrotron-induced exponential cutoff Lorentz factor,~$\theta_{\rm syn}$, analogous to that,~$\theta_{\rm ss}$, that follows from Thomson IC cooling. To arrive at the formula for~$\theta_{\rm syn}$, one simply replaces~$U_{\rm rad}$ in~(\ref{eq:thss}) with~$U_{\rm B}$ (i.e., changing the definition of~$\gamma_{\rm cool}$ in that equation), which yields~$\theta_{\rm syn}'/\theta_{\rm ss}' = U_{\rm rad}' / U_{\rm B}' \leq 100$. Synchrotron radiation thus introduces an effective particle acceleration cutoff,~$\gamma_{\rm cut}' = \theta_{\rm syn}'$, that is just a factor of~$100$ or so beyond~$\theta_{\rm ss}'$.

In the final steady state,~$\theta_{\rm ss}' = \theta_{\rm ss,f}' < \gamma_{\rm KN}'$, implying that~$\gamma_{\rm cut}' = \theta_{\rm syn}' \leq 100 \theta_{\rm ss,f}' < 100 \gamma_{\rm KN}'$. This places us in the special regime discussed in Section~\ref{sec:gcutlow}, in which the comoving final pair-loaded magnetization,~$\sigma_{\rm f}'$, is given by~(\ref{eq:sigtaugg}) and the comoving convergence time is just~$t_{\rm conv}' \simeq 70 \lambda_{\rm mfp}' / c = 70 L' / (c \tau_{\gamma\gamma}')$. Now, because~$L$ is transverse to the jet velocity, it does not transform between primed and unprimed frames. This, in turn, implies that~$\tau_{\gamma\gamma}' = n_{\rm rad}' \sigma_{\rm T} L' = n_{\rm rad} \sigma_{\rm T} L / \Gamma_{\rm j} = \tau_{\gamma\gamma} / \Gamma_{\rm j}$, since~$n_{\rm rad} = U_{\rm rad} / \epsilon_{\rm rad}$ transforms as~$n_{\rm rad}' = n_{\rm rad} / \Gamma_{\rm j}$. These transformation rules allow us to write~$t_{\mathrm{conv},i}' = 70 L / (\tau_i' c) = 70 \jetdist_0 / (\jetopen \tau_i' \Gamma_{\rm j})$. Thus, the locality condition,~$t_{\rm conv}' < t_{\times 2}' = 2\jetdist_0/c\Gamma_{\rm j}$, is satisfied provided that~$\tau_i' > \tau_{\rm crit}' \equiv 35/\jetopen \simeq 7$.

\citet{mehlhaff_etal_2021} estimate the comoving pair-production optical depth,~$\tau_i'$, presented by BLR or HDR photons evaluated across the jet width,~$L'= L = \jetdist_0 \theta_{\rm j}$, as~$\tau_{\rm BLR}' \simeq 3$ and~$\tau_{\rm HDR}' \simeq 20$, both independent of~$\Gamma_{\rm j}$ in the primed frame. Thus, the BLR photons are nearly optically thick enough~($\tau_{\rm BLR}'\sim \tau_{\rm crit}'$) for turbulence to converge before the macroscopic jet properties change much. For the case of HDR illumination, the optical depth is higher, so thorough convergence is more likely. In both cases, however, the optical depths are within order-unity factors of the critical one,~$\tau_{\rm crit}' \simeq 7$, necessary to uphold the locality condition, and so, given the crudeness of the present order-of-magnitude analysis, should perhaps just be regarded as close to this value. In this spirit, we conclude simply that~$t_{\rm conv} \sim t_{\times 2}$ for both BLR and HDR illumination. A good fraction (if not quite~100 percent) of the asymptotic pair count,~$n_{\rm f}$, is hence pumped into the plasma within~$t_{\times 2}$.

\subsection{Expected FSRQ pair yield}
\label{sec:yield}
The main goal here is to estimate the pair yield,~$n_{\rm f}/n_0$, that would be realized for fully equilibrated turbulence (once pair production is complete). Our approach is to estimate the final comoving plasma temperature,~$\theta_{\rm ss,f}'$, and magnetization,~$\sigma_{\rm f}'$, using results from Section~\ref{sec:fp}. We then combine these estimates with simple assumptions about the background jet magnetic field strength,~$B_0'$, and plasma number density,~$n_0'$, in order to extract the final pair yield from the formula,~$\sigma_{\rm f}' = B_{\rm f}'^2 / (16 \pi n_{\rm f}' \theta_{\rm ss,f}' m_e c^2)$.

As discussed in Section~\ref{sec:tconvtesc}, particle acceleration is cut off by synchrotron radiative cooling, placing us in the regime,~$\gamma_{\rm cut}'<100 \gamma_{\rm KN}'$. This means (Section~\ref{sec:gcutlow}) that~$\theta_{\rm ss,f}' \sim \gamma_{\rm KN}'/20$ and that~$\sigma_{\rm f}'$ follows from~(\ref{eq:sigtaugg}). Plugging~$\tau_{\rm BLR}' \simeq 3$ and~$\tau_{\rm HDR}' \simeq 20$ into~(\ref{eq:sigtaugg}) yields~$\sigma_{\rm f,BLR}' \simeq 2$ and~$\sigma_{\rm f,HDR}' \simeq 20$. Meanwhile, the comoving Klein-Nishina Lorentz factors,~$\gamma_{\mathrm{KN},i}'$ (needed to estimate~$\theta_{\rm ss,f}'$), follow from the characteristic seed-photon energies,~$\epsilon_i$, as~$\gamma_{\mathrm{KN},i}' = m_e c^2 / (4 \Gamma_{\rm j} \epsilon_i) = \gamma_{\mathrm{KN},i}/\Gamma_{\rm j}$. Using~$\epsilon_{\rm HDR} \simeq 0.3 \, \rm eV$ and~$\epsilon_{\rm BLR} \simeq 10 \, \rm eV$ \citep{mehlhaff_etal_2021}, we estimate~$\gamma_{\mathrm{KN,BLR}} \sim 1 \times 10^4$ and~$\gamma_{\mathrm{KN,HDR}} \simeq 4 \times 10^5$. 

We write the initial plasma density,~$n_0'$, in terms of the initial \textit{cold} magnetization (ignoring plasma pressure):~$\sigma_{\mathrm{c},0}' \equiv B_0'^2 / 4 \pi n_0' m_e c^2$. The pair yield can then be written as
\begin{align}
\frac{n_{\rm f}}{n_0} = \frac{n_{\rm f}'}{n_0'} &= \left( \frac{B_f'}{B_0'} \right)^2 \left( \frac{\sigma_{\mathrm{c},0}'}{\sigma_{\mathrm{f}}'} \right) \left( \frac{1}{4\theta_{\rm ss,f}'} \right) \notag \\
&\simeq \left( \frac{\sigma_{\mathrm{c},0}'}{\sigma_{\mathrm{f}}'} \right) \left( \frac{5}{\gamma_{\mathrm{KN}}'} \right) \, .
\label{eq:yield}
\end{align}
In the last step, we dropped the ratio of final-to-initial magnetic energy densities,~$B_{\rm f}'^2/B_0'^2$, since, provided the locality condition,~$t_{\rm conv} < t_{\times 2}$, is satisfied, the magnetic energy density remains approximately constant throughout the turbulent plasma evolution (Sections~\ref{sec:pic} and~\ref{sec:fp}).

At this point, the missing ingredient in~(\ref{eq:yield}) is the initial cold magnetization,~$\sigma_{\mathrm{c},0}'$. We phrase this magnetization in terms of that imprinted in the jet at its base,~$\sigma_{\rm b} = B_{\rm b}^2 / 4 \pi n_{\rm b} m_e c^2$, where~$B_{\rm b}$ and~$n_{\rm b}$ are, respectively, the jet-base magnetic field and plasma number densities. We consider the ideal case that the jet propagates without substantial dissipation or contamination (e.g., through shear-flow instabilities on the jet walls) from the ambient medium. Then, the magnetic field is mostly toroidal (i.e., wraps around the jet) and frozen into the bulk flow, with~$B(\jetdist) \theta_{\rm j} d = \rm const.$ so that~$B(\jetdist) \propto 1/d$ \citep{begelman_etal_1984}. Meanwhile, the plasma density decays as~$n(\jetdist) \propto 1/d^2$. Together, these two profiles imply that~$\sigma_{\rm c}$ is constant in~$\jetdist$. Identifying~$n_0' = n'(\jetdist_0) = n(\jetdist_0)/\Gamma_{\rm j}$ and~$B_0' = B'(\jetdist_0) = B(\jetdist_0)/\Gamma_{\rm j}$, we can write~$\sigma_{\mathrm{c},0}' = B_0'^2/4 \pi n_0' m_e c^2 = \sigma_{\rm b} / \Gamma_{\rm j}$.

Equation~(\ref{eq:yield}) can now be written as~$n_{\mathrm{f}}/n_0 \simeq 5 \sigma_{\mathrm{b}}/(\gamma_{\mathrm{KN}} \sigma_{\mathrm{f}}')$, which evaluates to
\begin{align}
    \left( \frac{n_{\rm f}}{n_0} \right)_{\rm BLR} \sim \frac{\sigma_{\mathrm{b}}}{10^4} \quad \mathrm{and} \quad \left( \frac{n_{\rm f}}{n_0} \right)_{\rm HDR} \sim \frac{\sigma_{\mathrm{b}}}{10^6} \,
\end{align}
independently of~$\Gamma_{\rm j}$. Hence, if the jet-base magnetization exceeds~$10^4$, a significant pair enrichment can be expected from BLR-illuminated turbulence; if it exceeds~$10^6$, pair enrichment is also possible in the case of HDR-illumination. This magnetization, in turn, depends on the detailed plasma physics of jet launch. For M87*, theoretical models based on magnetic reconnection \citep{kimura_etal_2022, chen_etal_2023, hakobyan_etal_2023} in the context of magnetically arrested accretion \citep{ripperda_etal_2022} predict~$\sigma_{\rm b}$ from~$10^{4}$ up to~$10^8$. All of these models predict~$\sigma_{\rm b}$ high enough for BLR-illuminated turbulence to pump significant new pairs into the jet; some give sufficient~$\sigma_{\rm b}$ for significant pair enrichment by HDR-illuminated turbulence.

\subsection{Discussion of FSRQ model}
\label{sec:fsrqsummary}
We have constructed an analytic model, benchmarked by our PIC (Section~\ref{sec:pic}) and FP (Section~\ref{sec:fp}) simulations, to determine whether turbulence in FSRQ jets can lead to strong \textit{in situ} pair enrichment. In this model, turbulence is triggered in an initially pristine plasma at a distance~$\jetdist_0$ in the jet from the central engine, as shown in Fig.~\ref{fig:jet}. Background radiation from the BLR or the HDR allows for IC scattering and pair production.

Our estimates consider important conditions that must be met if significant pair enrichment is to occur. The turbulent plasma must (Section~\ref{sec:tconvtesc}) converge to its final, thermal, pair-saturated state in a time,~$t_{\rm conv}$, shorter than both: (1) the time,~$\tesc$, for the turbulent plasma to escape the BLR or HDR background radiation, and (2) the time~$t_{\times 2}$ for the turbulence to probe jet regions with significantly different plasma properties. In addition (Section~\ref{sec:yield}), the asymptotic pair yield,~$n_{\rm f}/n_0$, from turbulent pair production for the given plasma parameters must be large.

In Section~\ref{sec:tconvtesc}, we restrict our attention to trigger points,~$\jetdist_0$, inside the outer edge,~$R_{\rm HDR}$ or~$R_{\rm BLR}$, of the radiation field, which ensures the ordering~$t_{\times 2} < t_{\rm esc}$. The locality condition,~$t_{\rm conv} < t_{\times 2}$, then becomes sufficient for the ejection condition,~$t_{\rm conv}<\tesc$. We showed that the locality condition holds provided that the comoving optical depth of the seed-photon background,~$\tau_{\gamma\gamma}'$, exceed~$\tau_{\rm crit}' \simeq 7$. The optical depths furnished by the HDR and BLR radiation fields are on par with this value. Thus, turbulence converges, at least nearly, while remaining in a uniform jet region, pumping out a good fraction of its asymptotic pair yield.

We estimate the asymptotic pair yields in Section~\ref{sec:yield}. In the ideal case that the jet does not substantially dissipate or entrain much plasma before turbulence is triggered, the input (cold) magnetization to turbulence is that,~$\sigma_{\rm b}$, embedded in the jet at its base. The expected pair yields are then~$(n_{\rm f}/n_0)_{\rm BLR} = \sigma_{\rm b} / 10^4$ and $(n_{\rm f}/n_0)_{\rm HDR} = \sigma_{\rm b} / 10^6$. A significant pair enrichment thus requires a generous jet-base magnetization, but not an impossible one in light of theoretical jet-launch models \citep[][though these are all specialized to the case of M87*, which is perhaps not a good FSRQ analog]{kimura_etal_2022, chen_etal_2023, hakobyan_etal_2023}.

We now return to consider one last criterion in deciding whether turbulence can load FSRQ jets with pairs. Namely, we ask whether the convergence time,~$t_{\rm conv}$, is faster than the expected duration,~$t_{\rm turb}$, of turbulence. Turbulence ceases once its free-energy source has been exhausted. In our simulations (Sections~\ref{sec:pic} and~\ref{sec:fp}), we consider driven turbulence, in which free energy can be pumped in indefinitely through the driving mechanism. However, in reality, the magnetic field itself may furnish the main source of free energy in FSRQ jets. In this case, a decaying picture of turbulence is more appropriate, wherein the magnetic energy dissipates in several eddy turnover (Alfvén/lightcrossing) times of the outer scale,~$L$: e.g.,~$t_{\rm turb} \sim 10 L/c$. The condition,~$t_{\rm conv}'<t_{\rm turb}'$, then reduces to~$\tau_{\gamma\gamma}' > 7$, incidentally the same criterion as that for~$t_{\rm conv}' < t_{\times 2}'$ (see Section~\ref{sec:tconvtesc} for the expression of~$t_{\rm conv}'$). Thus, no substantial new caveats to our results are introduced by the requirement~$t_{\rm conv} < t_{\rm turb}$.

The gamma-ray signatures of pair-thermalized turbulence in FSRQ jets follow straightforwardly from the equilibrium temperature,~$\theta_{\rm ss,f}' = \gamma_{\rm KN}'/20$, corresponding to mean-squared Lorentz factor~$\langle \gamma'^2 \rangle = 12 \theta_{\rm ss,f}'^2$. This translates to a mean observed photon energy (ignoring cosmological redshift) of~$\langle \epsilon_{\mathrm{obs}} \rangle = 4 \Gamma_{\rm j}^2 \langle \gamma'^2 \rangle \epsilon_{\rm rad} / 3 \sim \gamma_{\mathrm{KN}}^2 \epsilon_{\rm rad}/25$, which, upon plugging in~(\ref{eq:eradcases}), evaluates to~$\langle \epsilon_{\rm obs,BLR} \rangle \sim 70 \, \rm MeV$ and~$\langle \epsilon_{\rm obs,HDR} \rangle \sim 2 \, \rm GeV$. These characteristic energies are independent of~$\Gamma_{\rm j}$. They should be taken with a grain of salt, though, since, once turbulence shuts down or escapes the background radiation field, the thermal regulation mechanism fails. The characteristic radiative signatures of such turbulence may then be erased from the observable spectrum. The pairs loaded into the jet, on the other hand, would persist even beyond this point.

In all this discussion, we have considered the pair yield of turbulence when the starting plasma is already composed exclusively of pairs. However, the initial composition in the jet may also have a strong ion component. If the ions are relativistically magnetized, then they behave essentially as non-radiative positrons, and our main pair-plasma results apply up until the electron convergence time,~$t_{\rm conv}$. Beyond~$t_{\rm conv}$, the accumulated electrons and positrons become much cooler than the non-radiative ions. Under such circumstances, \citet{zhdankin_etal_2021} found that the heating ratio from ions to electrons/positrons grows without bound: as the electrons and positrons keep cooling, the ions soak up an ever-increasing fraction of the injected energy, leading to further lepton cooling. Thus, while the final pair yield -- established in large part prior to~$t_{\rm conv}$ -- might still be similar as to what we predict in this work, the final temperature might be much cooler than the one,~$\theta_{\rm ss,f} \sim \gamma_{\rm KN} / 20$, that we find.

It is more difficult to speculate about the case of trans- or non-relativistic ions. In this situation, the ions would initially control the plasma inertia and, hence, the Alfvén speed. They would also likely have a larger initial gyroradius compared to electrons, thus consuming a larger fraction of the injected energy \citep{zhdankin_etal_2019}. Whether and under what conditions adequate lepton acceleration could occur to ignite enough pair production for the leptons to claim control of the overall Alfvén speed and, hence, self-regulate, is unclear.

\section{Conclusions}
\label{sec:conclusions}
In this work, we study turbulence illuminated by an intense background radiation field, relevant to gamma-ray emitting regions in FSRQ jets. In this regime, high-energy particles cool down by IC scattering soft photons from the radiative background. Sufficiently energetic particles, with Lorentz factors~$\gamma>\gamma_{\rm KN}$, scatter photons in the discrete Klein-Nishina regime. The resulting gamma-ray quanta lie above pair-production threshold with the background and may hence be absorbed to produce new electron-positron pairs.

We argue (Section~\ref{sec:sketch}) that such pair production should feed back on turbulence. Turbulent particle acceleration initially outpaces radiative cooling. As a result, a prominent power-law tail develops in the particle energy distribution, with numerous particles at high-enough energies to emit photons above pair-production threshold. Absorption of these photons then begins to pump fresh pairs into the turbulent plasma, weighing it down (i.e., lowering~$\sigma$) and, hence, suppressing the efficiency of particle acceleration. This feedback terminates once enough pairs have accrued to limit particle acceleration to Lorentz factors below~$\gamma_{\rm KN}$. At this stage, particles radiate only in the low-energy Thomson IC regime, emitting solely photons beneath pair-production threshold and, hence, cutting off the supply of new pairs. In this late-time limit, radiative cooling is balanced against particle acceleration, leading to a thermal particle distribution at temperature~$\theta_{\rm f}$ \citep[as previously studied by][]{zhdankin_etal_2020}. This sequence of events represents a novel collisionless pair-production-mediated thermalization mechanism.

We then (Section~\ref{sec:pic}) use PIC simulations to verify that such pair thermalization indeed transpires. The PIC simulations are of radiative driven pair-plasma turbulence coupled to IC radiation and pair production as described above. They demonstrate clear convergence of the macroscopic plasma quantities to stable final values. This includes the predicted thermalization of the particle energy distribution to final temperature,~$\theta_{\rm f}$. A final pair-loaded magnetization,~$\sigma_{\rm f}$, is also reached; it is much lower than the value at the simulation onset (before turbulence ignites pair production).

In order to understand the main parameters governing the final plasma state, we supplement our PIC simulation with simpler and computationally cheaper Fokker-Planck (FP) models (Section~\ref{sec:fp}). Because of their lower cost, these FP simulations open up a broader region of parameter space than is accessible to PIC alone. They enable us to identify a strong dependence of the final temperature,~$\theta_{\rm f}$, final magnetization,~$\sigma_{\rm f}$, and time,~$t_{\rm conv}$, to converge to the final state, on~$\gamma_{\rm cut}/\gamma_{\rm KN}$, where~$\gamma_{\rm cut}$ is the particle acceleration cutoff Lorentz factor. In the limit,~$\gamma_{\rm cut} \leq 100 \gamma_{\rm KN}$, relevant to FSRQs, these three quantities obey the formulae,~$\theta_{\rm f} \simeq \gamma_{\rm KN}/20$,~$\sigma_{\rm f}^3/(1+\sigma_{\rm f}) \simeq (\tau_{\gamma\gamma}/2)^2$, and~$t_{\rm conv} \simeq 70 L / \tau_{\gamma\gamma} c$, where~$\tau_{\gamma\gamma}$ is the characteristic optical depth furnished by the background radiation over the turbulence driving scale,~$L$. Thus, pair-thermalization in FSRQ jets yields a final plasma state entirely independent of the initial plasma parameters; it is determined instead exclusively by properties of the background radiation (plus~$L$).

Finally, we perform back-of-the-envelope estimates (Section~\ref{sec:astroapps}) to determine whether pair-thermalized turbulence might boost the pair content of FSRQ jets \textit{in situ}. We find, under reasonable assumptions, that turbulence, once triggered in such jets, can converge: (1) while still in the presence of the strong ambient radiation impinging from the broad-line or hot-dust regions (BLR or HDR); (2) over a zone where the background jet properties remain approximately constant; and (3) before the (likely magnetic) free-energy source for turbulence is exhausted. We also estimate the pair yields for the separate cases of turbulence interacting with the BLR and HDR radiation fields, finding~$(n_{\rm f}/n_0)_{\rm BLR} \sim \sigma_{\rm b} / 10^4$ and~$(n_{\rm f}/n_0)_{\rm HDR} \sim \sigma_{\rm b} / 10^6$. The jet-base (cold) magnetizations,~$\sigma_{\rm b}$, needed for a substantial pair load are perhaps attainable in black-hole jets launched by the \citet{blandford_znajek_1977} mechanism \citep{kimura_etal_2022, chen_etal_2023, hakobyan_etal_2023}.

In general, \textit{what} blazar jets are made of and \textit{how} they accelerate particles are two outstanding questions in high-energy astrophysics. This study shows that these two issues, traditionally viewed as separate, may actually be deeply linked. As the jet propagates, the same dissipation sites that accelerate particles and power gamma-ray radiation may also enrich the jet with fresh electron-positron pairs. This study explicitly demonstrates this possibility in the context of turbulence (\citealt{mehlhaff_etal_2024} discuss it for the case of magnetic reconnection) in FSRQ jets. Although we have focused on pair plasmas, we conjecture (Section~\ref{sec:fsrqsummary}) that nearly the same pair loads could be achieved in the presence of an initial electron-ion plasma, as long as the ions are fairly relativistically magnetized. In this case, an initially electron-ion blazar jet could be transformed by \textit{in situ} turbulence into an electron-positron jet.

\begin{acknowledgments}
This project has received funding from the European Research Council (ERC) under the European Union’s Horizon 2020 research and innovation programme (grant agreement no.\ 863412). MZ acknowledges support from the Ambrose Monell Foundation and the Bezos Member Fund at the Institute for Advanced Study. VZ acknowledges support from NSF grant PHY-2409316. This work used Stampede2 and Stampede3 at the Texas Advanced Computer Center (TACC) and Purdue Anvil through allocation PHY160032 from the Advanced Cyberinfrastructure Coordination Ecosystem: Services \& Support (ACCESS) program, which is supported by U.S. National Science Foundation grants \#2138259, \#2138286, \#2138307, \#2137603, and \#2138296.
\end{acknowledgments}

\vspace{5mm}

\software{\textsc{Zeltron} \citep{cerutti_etal_2013, cerutti_werner_2019, mehlhaff_etal_2024},
          \textsc{NumPy} \citep{harris_etal_2020},
          \textsc{Matplotlib} \citep{hunter_2007}
}

\appendix
\section{Diffusive update in Fokker-Planck simulations}
\label{sec:fptech}
The Fokker-Planck (FP) simulations of Section~\ref{sec:fp} are predicated on evolving~(\ref{eq:fpfull}) in time. To do this, we employ a PIC-like numerical scheme. The distribution function~$f(\gamma, t)$ is represented by a number of delta-function-like samples. We call these samples particles, just like those of an ordinary PIC code, except that instead of living in six-dimensional phase space, they inhabit a one-dimensional energy space. Each particle~$i=1, \dots, N$ is identified only by its weight~$w_i$ and its Lorentz factor~$\gamma_i$. It is the particle Lorentz factors, rather than the distribution function itself, that are directly evolved (pushed) in time:~$\gamma_i = \gamma_i(t)$. The distribution function is simply reconstructed by binning the evolved particles onto a predefined grid of Lorentz-factors~$\tilde{\gamma}_n$. Adopting the convention that integer~$n$ corresponds to the bin centers, and half-integer~$n$ to the bin edges, we have~$f(\tilde{\gamma}_{n},t) = \Sigma_{\{i | \tilde{\gamma}_{n-1/2} < \gamma_i(t) < \tilde{\gamma}_{n+1/2}\}} w_i$.

Thanks to this particle description, the radiative part of the time evolution of~(\ref{eq:fpfull}) can be carried out in our FP simulations exactly as in our PIC simulations. The methods for this are described in detail by \citet{mehlhaff_etal_2024}. We adopt a Strang splitting to separate this radiative piece, which we conduct using preexisting methods, from the diffusive part of the update, which is new to this work. In what follows, we describe technical details of the diffusive part of our time evolution scheme, since we were not able to find this material in the literature.

To evolve the particles from time~$t=t_0$ to~$t=t_0 + \Delta t$, we give them random kicks in energy that are consistent with the diffusion-advection equation~(\ref{eq:fp}). To ensure this consistency, we derive an approximate (short-time) Green's function solution to~(\ref{eq:fp}). Each particle formally starts as a delta-function in energy space at time~$t_0$. At time~$t_0+\Delta t$, the slightly broadened Green's function tells us the probability distribution that a particle receives a kick of size~$\Delta \gamma$. In the following derivation of this Green's function, we drop the subscript~$i$.

We first introduce the new variable~$\xi(\gamma)$ such that~$d \xi / d \gamma 
 = 1 / \sqrt{D}$. This allows us to replace derivatives with respect to~$\gamma$ in equation~(\ref{eq:fp}) according to
 \begin{align}
     \partial_\xi = \frac{d \gamma}{d \xi} \partial_\gamma = \sqrt{D} \partial_\gamma \, .
 \end{align}
 Multiplying~(\ref{eq:fp}) by~$\sqrt{D}$, we have
 \begin{align}
     \partial_t \left( \sqrt{D} f \right) = \partial_\xi^2 \left( \sqrt{D} f \right) - \partial_\xi \left[ \left( \partial_\gamma \sqrt{D} + 2 \sqrt{D} / \gamma + A / \sqrt{D} \right) \sqrt{D} f \right] \, .
     \label{eq:fpmassaged}
 \end{align}
 Here, we have ignored the time dependence of~$D$, which, in the main text, is inherited through the dependence of~$t_{\rm acc}$ on the instantaneous value of~$\sigma$. We assume here that this time dependence, over the short interval~$\Delta t$, is negligible compared to the broadening of the initial delta-function,~$f(\gamma, t_0) = w \delta(\gamma - \gamma_0)$. Of course, for a truly time-independent~$D(\gamma)$, this argument is unnecessary.
 
 Introducing the function~$\mathcal{F} = \sqrt{D} f$ and the generalized advection coefficient~$\mathcal{A}(\xi) = d \sqrt{D} / d \gamma + 2 \sqrt{D} / \gamma + A / \sqrt{D}$, expression~(\ref{eq:fpmassaged}) becomes
 \begin{align}
     \partial_t \mathcal{F} = \partial_\xi^2 \mathcal{F} - \partial_\xi \left( \mathcal{A} \mathcal{F} \right) \, ,
     \label{eq:fpmassaged2}
 \end{align}
a diffusion-advection equation with diffusion coefficient equal to \textit{one}. We are thus tasked with finding the Green's function to this equivalent but simplified equation. The needed impulse initial condition, phrased in terms of~$\mathcal{F}$, is~$\mathcal{F}(\xi,t_0) = \mathcal{F}_0 \delta(\xi - \xi_0)$, where~$\xi_0 \equiv \xi(\gamma_0)$ and~$\mathcal{F}_0 = \sqrt{D} \, w \, (d\xi / d \gamma) = w$.

At this stage, it is convenient to make a \textit{short-time} hypothesis. Namely, because we are only evolving over a small time interval~$\Delta t$, we suppose that the initial distribution,~$\mathcal{F} = \mathcal{F}_0 \delta(\xi-\xi_0)$, does not have time to broaden very much. This means that~$\mathcal{A}$ does not change substantially across~$\mathcal{F}$, and we can sneakily replace it with its initial value,~$\mathcal{A}_0 = \mathcal{A}(\xi_0,t_0)$. Equation~(\ref{eq:fpmassaged}) then becomes
\begin{align}
    \partial_t \mathcal{F} = \partial_\xi^2 \mathcal{F} - \mathcal{A}_0 \partial_\xi \mathcal{F} \, .
\end{align}
The solution, starting from~$\mathcal{F}(\xi,t_0) = \mathcal{F}_0 \delta(\xi - \xi_0)$, is simply
\begin{align}
    \mathcal{F}(\xi, \Delta t) = \frac{\mathcal{F}_0}{\sqrt{4 \pi \Delta t}} \exp \left[ -\frac{1}{2}\left( \frac{\xi - (\xi_0 + \mathcal{A}_0 \Delta t)}{\sqrt{2 \Delta t}} \right)^2 \right] \, .
    \label{eq:greens}
\end{align}
This is just a normal distribution, initially centered at~$\xi_0$, that moves to the right at velocity~$\mathcal{A}_0$ and has broadening standard-deviation~$\sqrt{2 \Delta t}$.

Equation~(\ref{eq:greens}) is the probability that a particle, with initial~$\gamma_0 = \gamma(t_0)$ associated with~$\xi_0$, changes its energy to that,~$\gamma(t_0 + \Delta t)$, corresponding to~$\xi$ in time interval~$\Delta t$.
We can capture this probabilistic diffusion/advection by giving each particle's Lorentz factor a random kick, calculated as follows:
\begin{enumerate}
    \item Calculate~$\xi_0 = \xi(\gamma(t_0))$.
    \item Draw a random number~$\mathcal{R}$ from a unit normal distribution.
    \item Calculate~$\xi_{\rm new} = \xi_0 + \mathcal{R} \sqrt{2 \Delta t} + \mathcal{A}_0 \Delta t$.
    \item Invert the relation~$\xi(\gamma)$ to calculate~$\gamma(t_0+\Delta t)$ from~$\xi_{\rm new}$.
\end{enumerate}
The necessary inversion of~$\xi(\gamma)$ can be done explicitly in the short-time limit by Taylor expanding. Abbreviating~$\gamma_{\rm new} = \gamma(t_0+\Delta t)$ and~$D'(\gamma) = d D / d \gamma$, we have
\begin{align}
    \gamma_{\rm new} &\simeq \gamma_0 + \frac{d \gamma}{d \xi} (\xi_{\rm new} - \xi_0) + \frac{1}{2} \frac{d^2 \gamma}{d \xi^2} (\xi_{\rm new} - \xi_0)^2 \\
    &= \gamma_0 + \sqrt{D(\gamma_0)} (\xi_{\rm new} - \xi_0) + \frac{1}{4} D'(\gamma_0) (\xi_{\rm new} - \xi_0)^2 \, .
\end{align}
To arrive at the second line, we used~$d \gamma / d \xi = \sqrt{D}$ and
\begin{align}
    \frac{d^2 \gamma}{d \xi^2} = \frac{d}{d \xi} \sqrt{D} = \frac{1}{2 \sqrt{D}} \frac{d D}{d \xi} = \frac{1}{2 \sqrt{D}} \frac{d \gamma}{d \xi} \frac{d D}{d \gamma} = \frac{1}{2} \frac{d D}{d \gamma} \, .
\end{align}
Finally, plugging in~$\xi_{\rm new} - \xi_0 = \mathcal{R} \sqrt{2 \Delta t} + \mathcal{A}_0 \Delta t$, and using~$\mathcal{A}_0 = D'(\gamma_0) / [2 \sqrt{D(\gamma_0)}] + 2 \sqrt{D(\gamma_0)} / \gamma_0 + A(\gamma_0) / \sqrt{D(\gamma_0)}$, yields
\begin{align}
    \gamma_{\rm new} = \gamma_0 + \mathcal{R} \sqrt{2 \Delta t D(\gamma_0)} + \Delta t \left[ \frac{D'(\gamma_0)}{2} \left( 1 + \mathcal{R}^2 \right) + \frac{2 D(\gamma_0)}{\gamma_0} + A(\gamma_0) \right] + \mathcal{O}(\Delta t^{3/2}) \, .
    \label{eq:gnew}
\end{align}
This gives the simplified procedure to update a particle Lorentz factor from~$\gamma_0$ at time~$t_0$ to~$\gamma_{\rm new}$ at time~$t_0 + \Delta t$:
\begin{enumerate}
    \item Draw a random number~$\mathcal{R}$ from a unit normal distribution.
    \item Update~$\gamma_0$ to~$\gamma_{\rm new}$ by evaluating equation~(\ref{eq:gnew}).
\end{enumerate}

\bibliography{ref}
\bibliographystyle{aasjournal}

\end{document}